\documentclass[a4paper,fleqn,usenatbib]{mnras}

\usepackage{newtxtext,newtxmath}

\usepackage[T1]{fontenc}
\usepackage{ae,aecompl}

\usepackage{graphicx}	
\usepackage{amsmath}	
\usepackage{amssymb}	
\usepackage{longtable}
\usepackage[labelfont=bf]{caption}
\usepackage{lscape,supertabular}
\usepackage{natbib}

\title[The Non-Uniformity of Galaxy Cluster Metallicity Profiles]{The Non-Uniformity of Galaxy Cluster Metallicity Profiles}

\author[L. Lovisari et al.]{
L. Lovisari,$^{1}$\thanks{E-mail: lorenzo.lovisari@cfa.harvard.edu}
T. H. Reiprich,$^{2}$
\\
$^{1}$Harvard-Smithsonian Center for Astrophysics, 60 Garden Street, Cambridge, MA 02138, USA\\
$^{2}$Argelander Institute for Astronomy (AIfA), University of Bonn, Auf dem H\"ugel 71, 53121 Bonn, Germany\\
}

\date{Accepted XXX. Received YYY; in original form ZZZ}

\pubyear{2018}

\begin{document}
\label{firstpage}
\pagerange{\pageref{firstpage}--\pageref{lastpage}}
\maketitle

\begin{abstract}
We study a sample of 207 nearby galaxy groups and clusters observed with XMM-Newton. Key aspects of this sample include the large size, the high data quality, and the large diversity of cluster dynamical states. We determine the overall metallicity within 0.3R$_{500}$ and the radial distribution of the metals. On average, we find a mild dependence of the core metallicity with the average temperature of the system in agreement with previous results. However, we identify the cause of this mild dependence to be due to relaxed systems only; disturbed systems do not show this trend, on average. The large scatter observed in this relation is strongly associated with the dynamical state of the systems: relaxed systems have on average a higher metallicity in the core than disturbed objects. The radial profiles of relaxed systems are centrally peaked and show a steep decrease with radius, flattening beyond 0.3-0.4R$_{500}$. The metallicity of disturbed systems is also higher in the center but at much lower values than what is observed for relaxed objects. This finding is consistent with the picture that cluster mergers mix the abundance distribution by inducing large scale motions. The scatter of the radial profiles is quite large, but while for relaxed systems it decreases almost monotonically as function of the radius, for disturbed systems it shows a significant boost at large radii. Systems with a central radio source have a flatter profile indicating that central AGNs are an efficient mechanism to uplift and redistribute the metals in the ICM.
\end{abstract}

\begin{keywords}
X-rays: galaxies: clusters -- galaxies: clusters: intracluster medium -- galaxies: groups: general\end{keywords}



\section{Introduction}
Groups and clusters of galaxies are unique environments for elemental
abundance measurements because, thanks to their deep potential, they
are thought to retain all the metals that have been produced over
cosmic time. The distribution of the metals in the ICM depends on the star formation history (and so the number of supernovae exploded so far) and on the efficiency of the different processes that injected the enriched material into  the ICM (\citealt{2008SSRv..134..363S}, and references therein). 
Thus, the study of the abundance and distribution of heavy
elements in the ICM provides vital information on the mechanisms
responsible for the metal transfer from galaxies to the hot gas, and
on the different roles played by mergers, galactic winds, ram-pressure
stripping and AGN outbursts (e.g. \citealt{1997ApJ...488...35R}, \citealt{2009MNRAS.399..239R},
\citealt{2010MNRAS.401.1670F}, \citealt{2014A&A...569A..31H}, \citealt{2018arXiv181101955B}).
X-ray spectra provide an accurate measurement of the metallicity of the intracluster medium (ICM).  Since the first X-ray observations of the iron line feature in the 1970s by \cite{1977MNRAS.178P..75M}, observations
have confirmed that the ICM contains both primordial elements as well
as heavy elements  (e.g. \citealt{2004A&A...419....7D,2009A&A...508..565D}). However, metals are not distributed uniformly in the ICM as shown by the studies of the metallicity spatial distribution (e.g. \citealt{2009A&A...493..409S}, \citealt{2009A&A...508..191L,2011A&A...528A..60L}).

For many years it was almost impossible to perform metallicity
measurements to large fractions of $R_{500}$, but with XMM-Newton, Chandra, and
Suzaku\footnote{Because of the malfunction experienced by Suzaku in
  early June 2015, the mission has ended.}, it has
been possible to derive metallicity profiles for a sizable  sample of
galaxy groups and clusters (e.g. see the review by \citealt{2018arXiv181101967M}). These azimuthally averaged profiles show a centrally
peaked metallicity distribution in cool-core systems and a relatively
flat distribution in non-cool core clusters (e.g. \citealt{2004A&A...419....7D}, \citealt{2007A&A...461...71P}, \citealt{2008A&A...487..461L}). Outside the
cluster core, and irrespective of their central properties, the
profiles gently decline out to $\sim$0.3R$_{500}$. Beyond that radius,
the metallicity is consistent with being flat, as shown by the measurements of the ICM metallicity up to the virial radius for several clusters (e.g. \citealt{2008PASJ...60S.343F},  \citealt{2013Natur.502..656W}, \citealt{2015ApJ...811L..25S}, \citealt{2016A&A...592A..37T}, \citealt{2017ApJ...836..110E}, \citealt{2017MNRAS.470.4583U}).

A proper modeling of the temperature is crucial, otherwise the combination of different temperatures, either due to a strong gradient or to a multiphase plasma, can lead to important biases. \cite{2000MNRAS.311..176B} showed that fitting a multi-temperature plasma with a single temperature model leads to an underestimation of the metallicity. This effect is known as ``iron bias''. \cite{2008ApJ...674..728R} by analyzing mock XMM-Newton observations of simulated galaxy clusters, showed that projection effects, low spatial resolution, and a particular temperature range (i.e. 2-3 keV) can also lead to a systematic overestimate of the metallicity, known as ``inverse iron bias'' (see also \citealt{2009A&A...493..409S} and \citealt{2010A&A...522A..34G}). 

The effect of these two biases can be an explanation for the trend found by \cite{2005ApJ...620..680B} in the abundance-temperature plot. Using the ASCA archive observations of 273 galaxy clusters they found a constant abundance at a value of 0.3 solar for high temperature clusters, an averaged  high metallicity (larger by up to a factor
of 3 than hotter systems) for the objects in the 2-4 keV energy range, and finally an abundance drop for low temperature systems.  
Alternatively, this trend could indicate that different enrichment mechanisms play different roles in clusters with
different masses, or that the star formation is more efficient in smaller clusters, as suggested also by
optical and near-infrared observations of nearby systems (\citealt{2003ApJ...591..749L}; \citealt{2007ApJ...666..147G}). However, it is hard to conceive a mechanisms at work only in clusters with a temperature in the 2-4 keV range. It is easier to focus on some possible measurement bias, which has to bias iron abundances high. 

A step forward in our understanding of the radial distribution of the abundances in giant elliptical galaxies, groups, and clusters of galaxies was done by the CHEERS (CHEmical Enrichment RGS Sample, see \citealt{2017A&A...607A..98D}) collaboration. Analyzing the 44 CHEERS systems \cite{2017A&A...603A..80M} found a significant negative radial metallicity gradient out to 0.9$R_{500}$ for hot systems (k$T$$>$1.7 keV) and 0.6$R_{500}$ for cool systems (k$T$$<$1.7 keV) with the latter having on average a lower metallicity than the most massive clusters.  By construction the CHEERS sample of 44 objects contains basically only cool-core systems and is dominated by galaxy groups (i.e. average kT$\lesssim$3 keV), so it is not representative of the whole cluster population. Other recent studies (\citealt{2015A&A...578A..46E}, \citealt{2016ApJ...826..124M}, \citealt{2017MNRAS.472.2877M}), although more focused on the evolution of the metallicity with redshift, found a trend of higher metallicity in the innermost regions of cool-core clusters with respect to non-cool-core systems. Moreover, the metallicity in the core was found to be much higher than the metallicity in the outskirts for both relaxed and disturbed systems (e.g. \citealt{2015A&A...578A..46E}, \citealt{2017MNRAS.472.2877M}). While these studies are based on more representative galaxy clusters samples for most of the systems they could not derive very fine spatial binning because of the high cluster redshifts which lead to an average low number of counts per cluster.

In this paper we investigate the metallicity for a large sample of nearby clusters for which we can derive fine spatial binning and extend the study of the radial metallicity profiles to a larger sample of hot clusters with respect to what has been done with the CHEERS sample. Since a systematic comparison between relaxed and disturbed systems can help to shed light into the mixing of the metals in the ICM, we investigate the impact of the dynamical state in the metallicity profiles of galaxy groups and clusters.  Moreover, we investigate the impact of the central AGNs in the distribution of the central abundances. 

Throughout the paper we assume a $\Lambda$CDM cosmology with $H_0$=70 km/s/Mpc, $\Omega_{\Lambda}$=0.7 and $\Omega_m$=0.3. The estimated metallicities are all relative to the solar values given by \cite{2009ARA&A..47..481A}. The outline of the paper is as follows. The data preparation and analysis is presented in \S2 and the results in \S3. In \S4 and \S5 we discuss the results and present our conclusions. 

\section{Observations and data preparation}
\subsection{Sample selection}
The aim of this work is the study of the metal enrichment of the ICM in groups and clusters of galaxies, and the determination of the impact of merging and interactions in the distribution of the metallicity in the ICM.  We choose to look at a sample of local galaxy groups and clusters with available XMM-Newton data. 
As input catalog we used the MCXC catalogue (see \citealt{2011A&A...534A.109P}) and restricted the search to all the objects in the NORAS, REFLEX and eBCS sub-catalogs (\citealt{2000ApJS..129..435B}, \citealt{2004A&A...425..367B}, \citealt{1998MNRAS.301..881E}, and \citealt{2000MNRAS.318..333E}). By arbitrary using an upper redshift cut of 0.1 we found  207 objects with useful (i.e. not flared) XMM-Newton observations (the list of objects can be found in Appendix of this paper). Clusters with multiple X-ray peaks, and clusters with an estimated 0.3$R_{500}$\footnote{As reference for  $R_{500}$ we used the values provided by \cite{2011A&A...534A.109P}} extending beyond the XMM-Newton field-of-view (FOV) were removed from the sample. 

\subsection{Data reduction}
Observation data files (ODFs) were downloaded from the XMM-Newton archive and processed with the XMMSAS v16.0.0 software for data reduction. The initial data processing to generate calibrated event files from raw data was done by running the tasks {\it emchain} and {\it epchain}. We only considered single, double, triple, and quadruple events for MOS (i.e. PATTERN$\le$12) and single for pn (i.e.  PATTERN==0) and we applied the standard procedures for bright pixels and hot columns removal (i.e. FLAG==0) and pn out-of-time correction.
All the data sets were cleaned for periods of high background due to the soft protons following the two stage filtering process procedure extensively described in \citet{2011A&A...528A..60L}. Briefly, a light
curve was first extracted in 100 s bins in  the 10-12 (12-14) keV energy band for MOS (pn).   A Poisson distribution was fitted to the histogram of this light curve, and $\pm$2$\sigma$ thresholds calculated.
The Good Time Interval (GTI) files were produced using these thresholds,  and  the  event  lists  were filtered  accordingly.  The new event lists were then re-filtered in a second pass as a safety check for possible flares with soft spectra (e.g., \citealt{2004A&A...419..837D}; \citealt{2005ApJ...629..172N}; \citealt{2005A&A...443..721P}). In this case, light curves were made with 10s bins in the full [0.3-10] keV band. The point-like sources were detected with the {\it edetect-chain} task and excluded from the event files. The background event files were cleaned by applying the same PATTERN selection, flare rejection, and point-source removal as for the corresponding target observations.

\subsection{Background treatment}
The total background consists of many different components, each one characterized by distinct temporal, spectral, and spatial variations (see the Table\footnote{\url{https://www.cosmos.esa.int/web/xmm-newton/epic-background-components}} summarizing the different components at the XMM-Newton background analysis webpage). The main components are the non-vignetted quiescent particle background (QPB) and the cosmic X-ray background (CXB).
The CXB can be then subdivided into three subcomponents: thermal emission from
the Local Hot Bubble (LHB) and from the Galactic Halo (GH), and an extragalactic component representing the unresolved emission from AGNs. 

Since it is the result of the emission of astrophysical sources the CXB is folded with the response files. To model this component we followed the method presented in \citet{2008A&A...478..615S}. Basically, the XMM-Newton spectra were fitted simultaneously with the ROSAT All-Sky Survey (RASS) spectra extracted from the region just beyond the virial radius using the available tool\footnote{\url{http://heasarc.gsfc.nasa.gov/cgi-bin/Tools/xraybg/xraybg.pl}} at the HEASARC webpage. Both the LHB and GH were described by a thermal emission with temperatures free to vary, and  with metallicity and redshift frozen to 1 and 0, respectively. The GH component is absorbed by a gas with total (i.e. neutral and molecular, see appendix A of this manuscript and \citealt{2013MNRAS.431..394W} for more details) hydrogen column density estimated using an online tool\footnote{\url{http://www.swift.ac.uk/analysis/nhtot/index.php}}, while the LHB is not. The emission to account for the unresolved point sources was modeled with an absorbed power-law with its slope set to 1.41 \citep{2004A&A...419..837D}.

The QPB consists of a continuum component and several fluorescent lines (see e.g. \citealt{2015A&A...575A..37M} for a list of those lines) which vary across the detector. The filter wheel closed (FWC) observations can be used to estimate the intensity of the various components. We first renormalized the FWC observations following the procedure explained in \citet{2009ApJ...699.1178Z}. Then for every interesting region (e.g. the different annuli of the radial profiles) we extracted a spectrum from the same detector area and we fitted it with a broken power-law\footnote{The first power-law component account for the QPB continuum while the second for a strong low-energy tail due to the intrinsic noise of the detectors.} plus 8 (9) gaussian lines for MOS (pn). We then included such a model in the fit with the normalizations of the broken power-law free to vary within a $\pm3\%$ to account for  the uncertainties associated with the normalization factor used to rescale the FWC observations which are on the order of 3-5$\%$ for relatively short observations.

Finally, we added an extra power-law, folded only with the RMF\footnote{Ideally, since the protons are funneled toward the detectors by the X-ray mirrors, the power-law should be folded also through the ARF. However, the proton vignetting is different from the photon vignetting (e.g. \citealt {2017ExA....44..297M}) and to date, there is not proton vignetting model available.}, to the background modeling to account for a residual soft proton contamination which is affecting many observations even after filtering soft flare events. Following the suggestion of  the ESAS coobook{\footnote{\url{ftp://xmm.esac.esa.int/pub/xmm-esas/xmm-esas.pdf}}} we allowed the spectral index of this component free to vary in the range 0.1-1.4.  Since this component may be different for MOS and pn detectors, both the slope and normalization are left free to vary in the three detectors and in all the regions of interest (this account, in first approximation, for the proton vignetting).

\subsection{Spectral analyses}
Our goal is to determine both an overall cluster core abundance and, when the data quality allow, the abundance profiles. The objects in our sample span a large range of masses and redshifts, so that their extension in the sky is very different. To have a fair comparison between the core abundances in different systems we extract spectra from a region within 0.3$R_{500}$.  This choice allows us to have a high  S/N=S/$\sqrt{S+N}$ and to focus on the radial region where groups and clusters show a metallicity enhancement. 

All the regions used for the abundance profiles were centered on the peak of the X-ray emission.  
The size of the annuli  have been determined by requiring a minimum width of 30$^{\prime\prime}$ and a fixed S/N. The first requirement ensures that most of the flux (i.e. $>80\%$, see \citealt{2009ApJ...699.1178Z}) comes from the selected region (due to the XMM-Newton PSF some photons scatter from one annulus to another), the second that the abundances are determined with a relatively low and homogeneous uncertainty.  Due to the fading of the emission lines for increasing temperatures, hotter systems require a higher S/N to determine the abundances with a similar uncertainty of galaxy groups. Thus, the S/N in the 0.3-10 keV band used for different clusters was based on their overall  core temperature (i.e. $<0.3$R$_{500}$).  The background level for the calculation of the S/N was determined using the FWC observations for the particle background component and using the results from \cite{2002A&A...389...93L}  for the foreground component and rescaled to the area of interest.
In Appendix B we give more details about our choice.  \\
All the extracted spectra were re-binned to ensure at least 25 counts per bin which is necessary for the validity of the $\chi^2$ minimization method\footnote{ Because of the units (i.e. rates instead of counts) of the RASS spectrum (extracted with the HEASARC background tool v2.5.1)  which is jointly fitted with our XMM-Newton spectra we cannot use the cstat statistic in XSPEC which requires an integer number of counts per bin. We show in Appendix C that this is not biasing our results.}. Some clusters are very bright and our requirement of a minimum width for selecting the spectral regions leads to a large number of counts in some spectra (in particular in the central regions). In that case a statistical grouping can dramatically oversample the instrument resolution and cause problems during the spectral fitting, so using the SAS task {\it specgroup} we set the minimum energy width of each group to be at least 1/3 of the full width half maximum (FWHM) resolution at the central photon energy of the group.  \\
The spectral analysis was done using XSPEC (\citealt{1996ASPC..101...17A}). Spectra were fit in the full (i.e. 0.3-10 keV) energy band with two absorbed (using the total $N_H$) APEC thermal plasma model (\citealt{2001ApJ...556L..91S}) linking the abundances of the two gas phases, as the current instruments do not allow to measure the abundances separately and accurately. In each annulus, the  MOS and pn spectra were fitted simultaneously, with linked temperatures and abundances, and all normalizations free to vary to account for the calibration offsets between the different detectors (e.g. \citealt{2017AJ....153....2M}). Beyond $\sim$2 arcmin the second temperature component is basically unconstrained, even in presence of relatively good data. Thus, the outer annuli have been fitted with a single APEC model. 

\begin{figure}
	\includegraphics[width=\columnwidth]{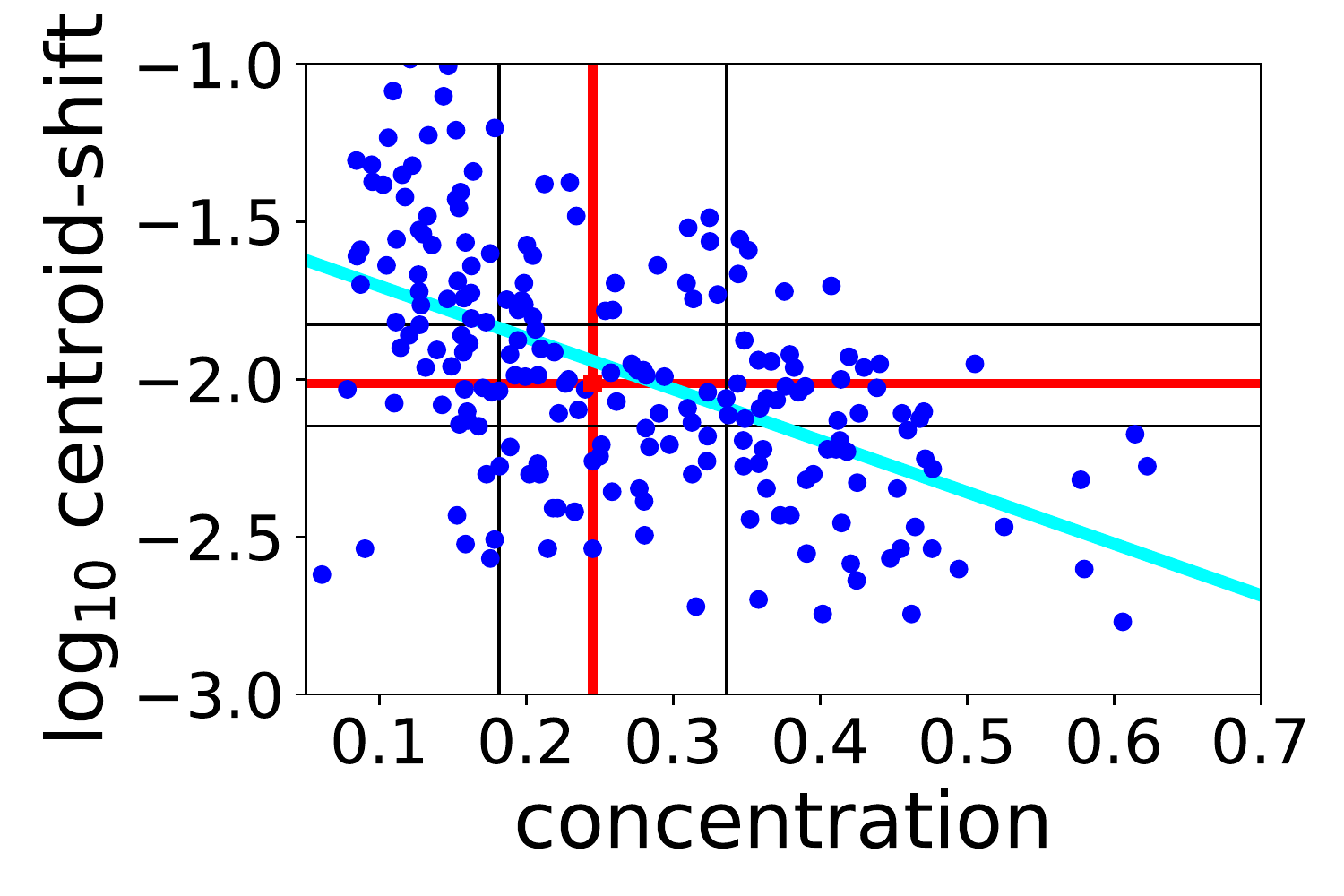}
    \caption{Concentration-centroid shift diagram.   The red lines indicate the median values while the black lines indicate the 33 and 67 percentile. The cyan line represents the best fit relation to the data points. For the centroid-shift values we plotted the decimal logarithm.}
    \label{fig:morpho}
\end{figure}

\subsection{Morphology} 
One goal is to compare the distribution of the metals between relaxed and disturbed systems. \cite{2017ApJ...846...51L} found that centroid-shift (\citealt{1993ApJ...413..492M}) and concentration (e.g. \citealt{2008A&A...483...35S}) are efficient parameters to classify the X-ray cluster morphology. For each cluster we computed these two parameters within 0.5$R_{500}$. We did not consider larger radii because most of the nearby objects lack  XMM-Newton coverage in the outer regions. The result of the morphological analysis is shown in Fig. 1. By fitting\footnote{Throughout the paper the fits
were performed using the  Bayesian code by \cite{2007ApJ...665.1489K}.} the individual points with a single power-law we get a slope of $-$1.64$\pm$0.18 with an intrinsic scatter of 0.33$\pm$0.11. Thus,  as expected, there is a clear anti-correlation between the two parameters, as also indicated by the Spearman rank coefficient $r$=$-$0.54 ($p$$<$0.01). The most relaxed objects are the one in the bottom-right quadrant while the most disturbed are in the top-left quadrant. Systems in the top-right panel have a concentrated surface brightness with substantial substructures at larger radii or strong ellipticity and can be considered as objects undergoing a minor merger that did not destroy the core but creates an inhomogeneous distribution of the ICM. Systems in the bottom-left panel can be interpreted as post-merger clusters where enough time to erase most of the inhomogeneities has passed,  but not sufficient to rebuild the peaked core.

\section{Results}
\subsection{Core metallicities}
In Fig. 2 we present the average metallicity, $A_{core}$, determined within 0.3$R_{500}$  as a function of the cluster temperature as determined within the same radius (k$T_{core}$).  A linear fit to the data gives a mild slope of $-$0.016$\pm$0.005 (scatter=0.121$\pm$0.040) indicating on average lower metallicity in the cores of the most massive systems. This result is confirmed by the median values (shown as green crosses) computed for different temperature intervals, and by the Spearman rank test which returned a value for the temperature-abundance correlation of r=$-$0.18 (p$<$0.01). While there are a few groups with very high central metallicity the mild correlation is confirmed even when excluding systems with k$T$$<$1 keV (slope=$-$0.015$\pm$0.005, r=$-$0.17 and p=0.02) or with k$T$$<$2 keV (slope=$-$0.016$\pm$0.005, r=$-$0.23 and p$<$0.01). Also, there is a clear increasing scatter in the low-mass regime, but we note that our sample has  a relatively low number of very  massive clusters since these are intrinsically rare systems.

\begin{figure}
\includegraphics[width=\columnwidth]{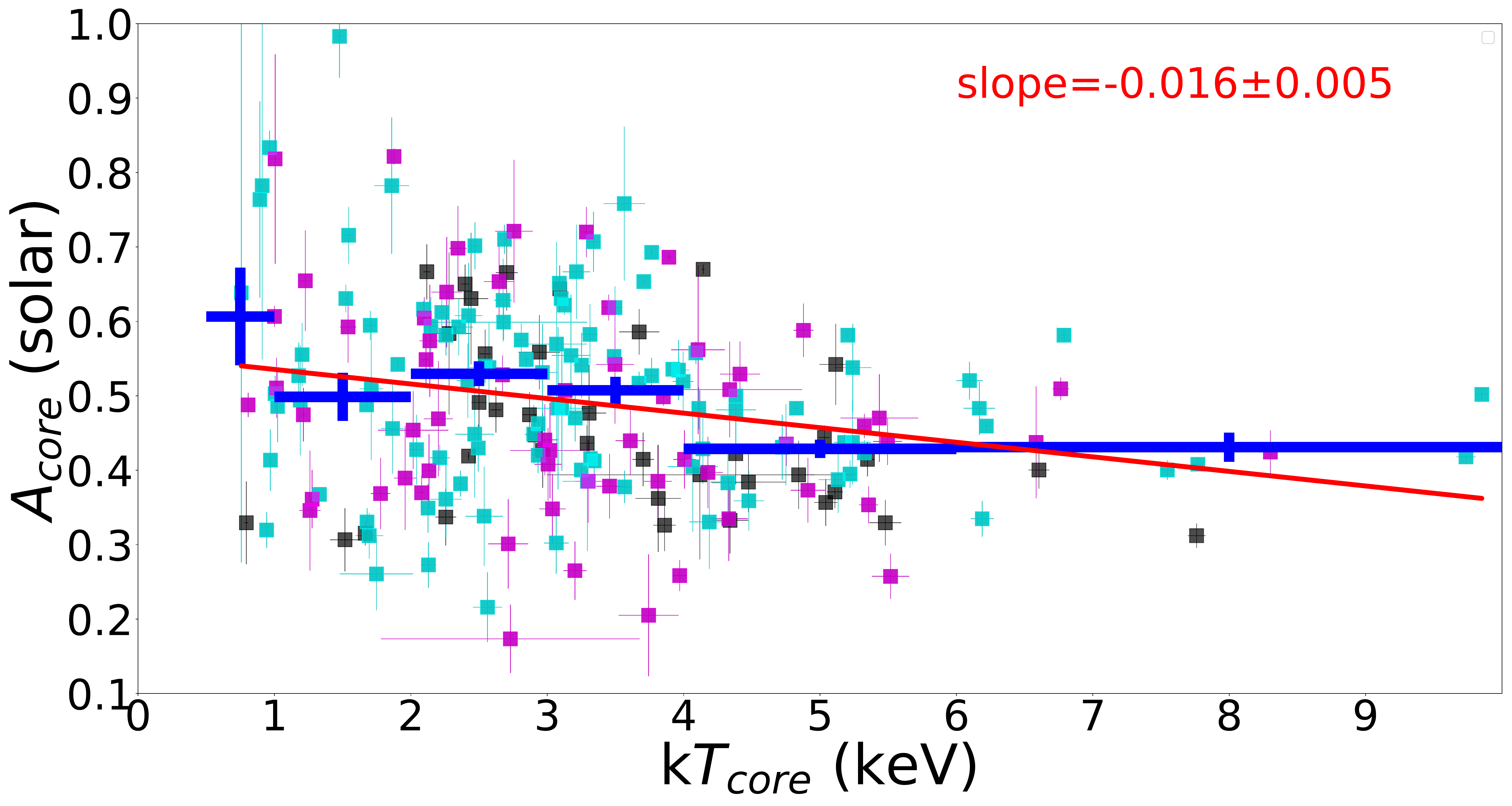}
\caption{Metallicity vs temperature plot obtained fitting the spectra extracted within 0.3R$_{500}$. The square points represent the single objects, while the blue crosses are the median values in each temperature bin with the error bars in the Y-axis representing the standard errors. Clusters with and without a central radio source are shown in cyan and magenta, respectively. In black we show the clusters which are not covered by the NVSS catalog.  The solid red line represents the best fit to all the unbinned data.}
\end{figure}

\begin{figure}
\includegraphics[width=\columnwidth]{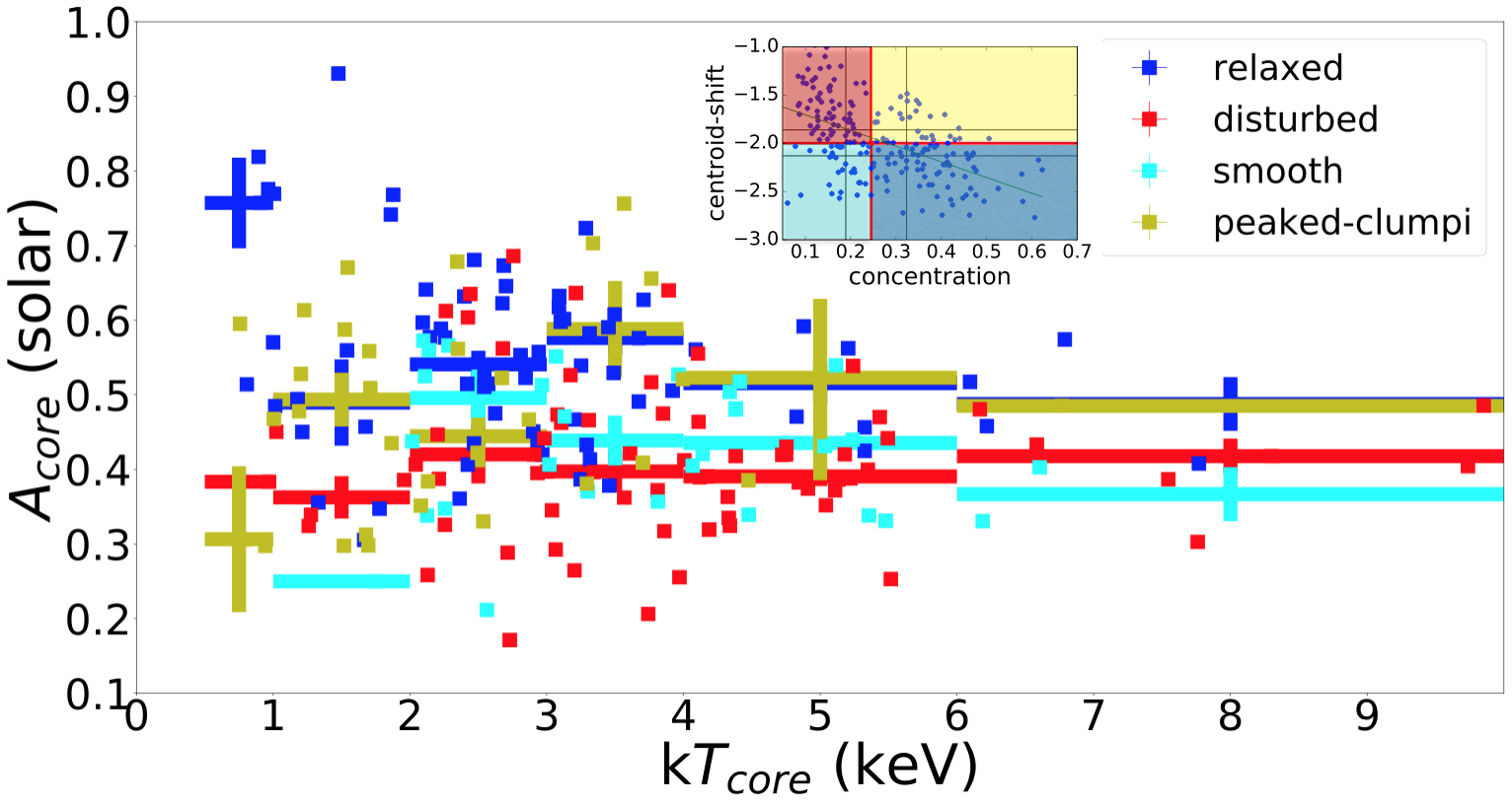}
\caption{The same as in Fig. 2 but highlighting with different colors clusters in the different quadrants of the concentration-centroid shift diagram.  We used the medians of concentration and centroid-shift distributions to split the sample. }
\end{figure}

\begin{figure*}
\includegraphics[width=\textwidth]{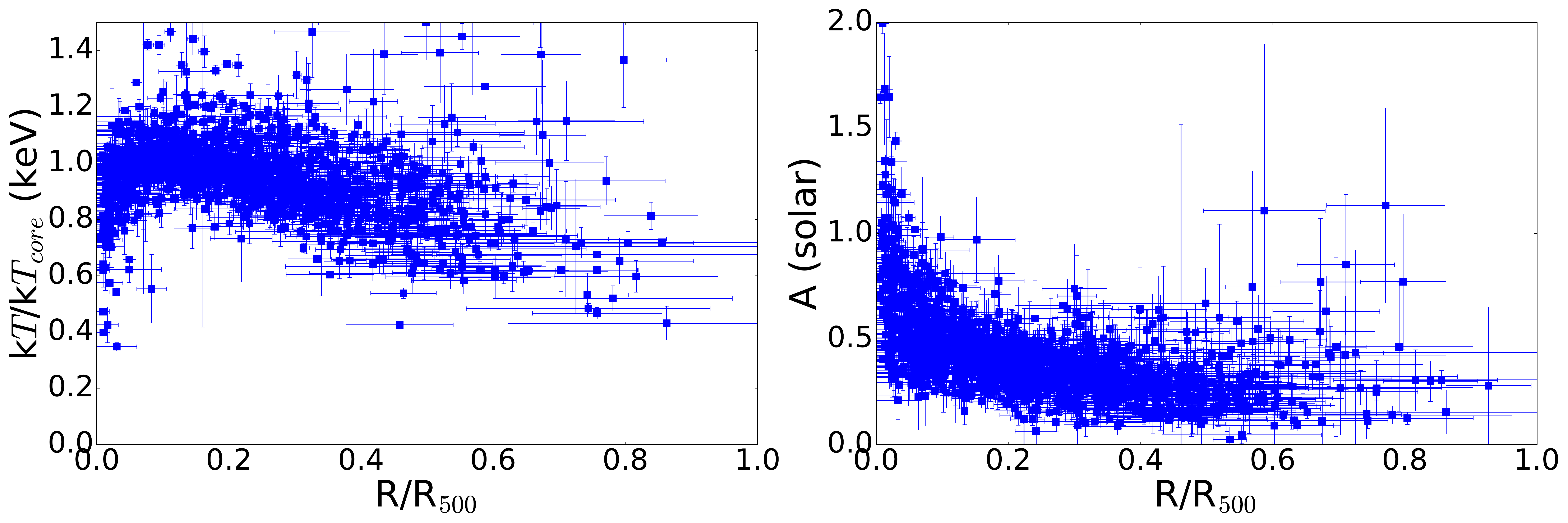}
\caption{Temperature ({\it left}) and abundance ({\it right}) profiles scaled by $R_{500}$. The temperature profiles have been also normalized by the average temperature measured within 0.3$R_{500}$.  }
\end{figure*}

\begin{figure*}
\includegraphics[width=\textwidth]{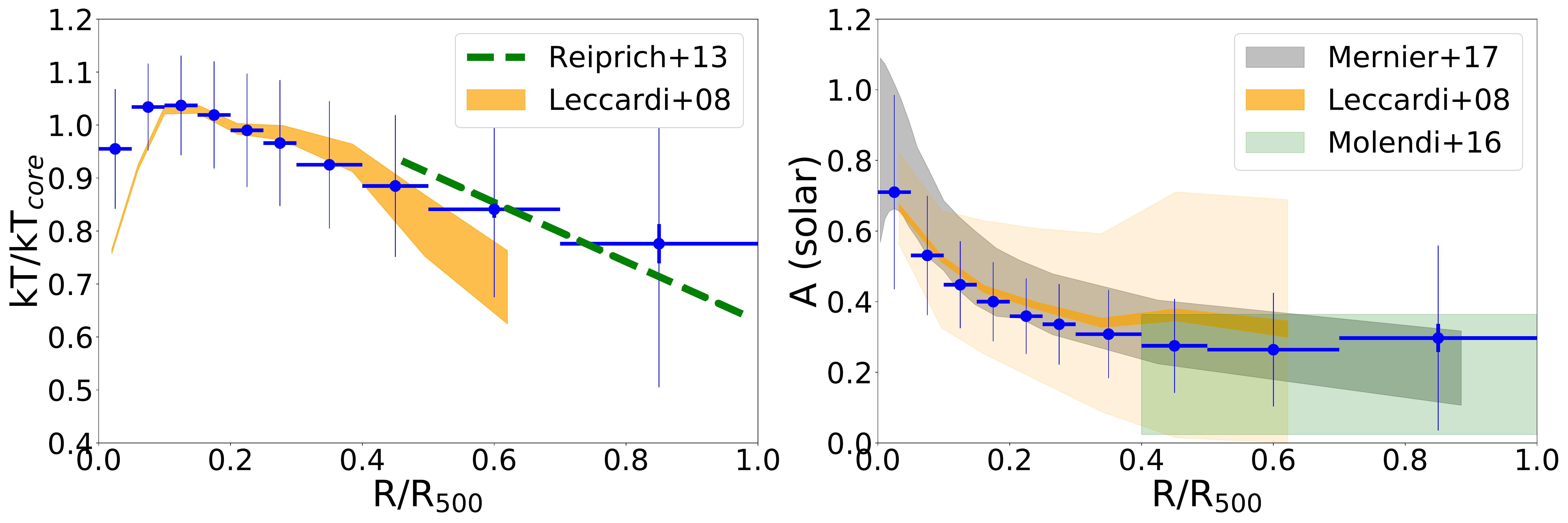}
\caption{Median temperature ({\it left}) and abundance ({\it right}) profiles scaled by $R_{500}$. The temperature profiles have been also normalized by the average temperature measured within 0.3$R_{500}$. Data points and errorbars show the median values and the scatter of the measurements in each radial bin. The statistical errors are shown as a thicker errorbar (smaller than the symbol size for most data points).  The grey shadow area represents the average profile, including the scatter, derived by \citealt{2017A&A...603A..80M}. The orange area represents the average profiles obtained by  \citealt{2008A&A...487..461L} with the dark and light colors illustrating the statistical uncertainties and the scatter, respectively. The green dotted line in the left panel is the best linear fit derived by \citealt{2013SSRv..177..195R} in the galaxy clusters outskirts.  The green area in the right panel shows the best estimate of the metallicity beyond 0.4$R_{500}$ derived by \citealt{2016A&A...586A..32M}.}
\end{figure*}

Relaxed and disturbed clusters may not be sharing the same average abundance in the studied volume. If so, their different fractions in different samples have an impact in the result shown in Fig. 2 and may lead to a wrong interpretation. Using the concentration and centroid-shift medians we subdivided the sample in  relaxed clusters (i.e. high concentration and low centroid-shift, bottom-right quadrant in Fig. 1) and disturbed (i.e. low concentration and high centroid-shift,  top-left quadrant in Fig. 1) clusters. We also show clusters that are classified as relaxed using the concentration and disturbed with the centroid-shift (top-right quadrant) and relaxed using the centroid-shift and disturbed by the concentration (bottom-left quadrant). The results are shown in Fig. 3. Apart from the innermost bin, the metallicities of the peaked-clumpi clusters are consistent with the ones of the relaxed systems and the metallicities of the smooth clusters are very similar with the ones of the disturbed systems. While this may be an indication that the driver for the high metallicity is the presence of a cool-core, we note  that the metallicity distribution that we obtain by splitting the sample using only the concentration is very similar to the one we obtain by splitting the sample using only the centroid-shift. 

Keeping in mind that there is no strong boundary between relaxed and disturbed systems and that the morphological parameters have some scatter due to projection effects, the result shows a very clear trend: the relaxed clusters have on average a higher metallicity in the core than the disturbed systems at all temperatures.  Moreover, by fitting relaxed and disturbed systems independently the abundance-temperature correlation disappears for dynamically active clusters. The Spearman rank test gives a probability p=0.95 of no correlation for the disturbed systems. Relaxed clusters still show the mild relation (slope=$-$0.022$\pm$0.010, r=$-$0.19 and p=0.11) but it becomes more uncertain when excluding kT$<1$keV  (slope=$-$0.014$\pm$0.011, r=$-$0.09 and p=0.48)  or  kT$<2$keV  (slope=$-$0.016$\pm$0.008, r=$-$0.20 and p=0.15)  systems.

\begin{figure*}
\includegraphics[width=\textwidth]{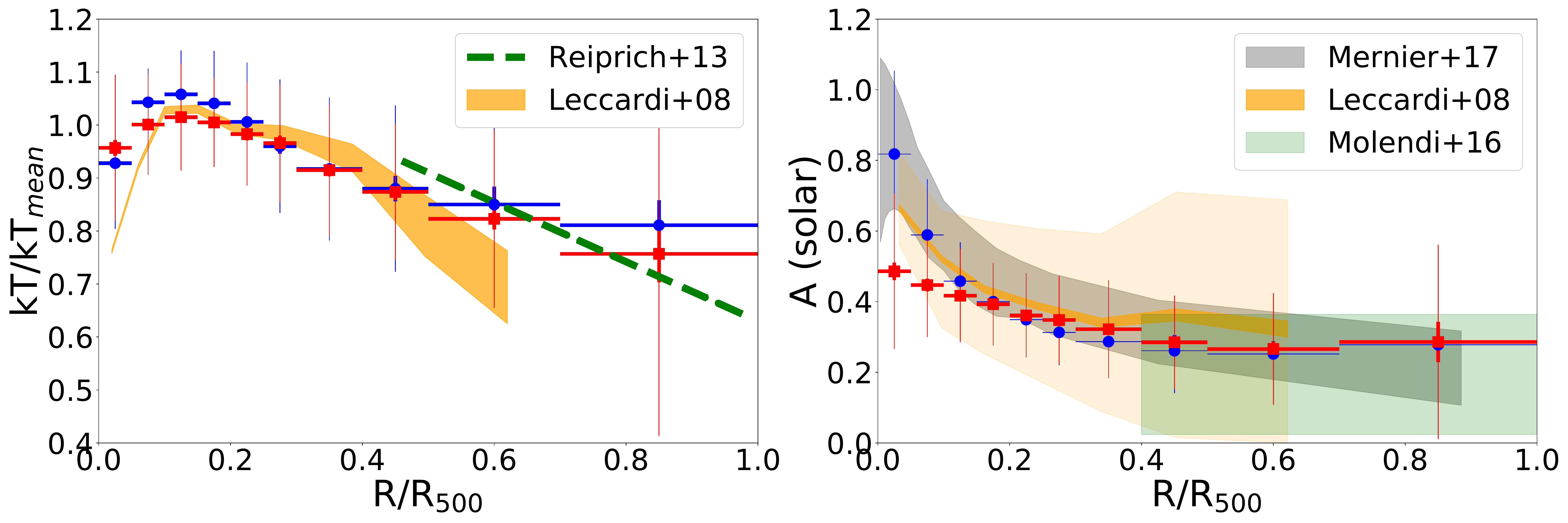}
\caption{Stacked temperature ({\it left}) and metallicity ({\it right}) profiles for relaxed (blue points) and disturbed (red squares) systems, using the median value of the concentration parameter to classify the systems.}
\end{figure*}

\begin{table}
\renewcommand{\thetable}{\arabic{table}}
\centering
\caption{Average metallicity profiles for relaxed and disturbed galaxy groups and clusters.}
\begin{tabular}{|c|cc|cc|}
\hline
\multicolumn{1}{|c|}{} & \multicolumn{2}{c|}{Relaxed} & \multicolumn{2}{c|}{Disturbed}  \\
\hline
Radius & A   & scatter  & A  & scatter \\
(r/r$_{500}$) & (solar) & &  (solar)  &\\
\hline
0.00-0.05 & 0.818$\pm$0.015 & 0.236$\pm$0.009  & 0.486$\pm$0.025   & 0.220$\pm$0.011    \\
0.05-0.10 & 0.589$\pm$0.014 & 0.158$\pm$0.006  & 0.447$\pm$0.018   & 0.147$\pm$0.015    \\
0.10-0.15 & 0.458$\pm$0.011 & 0.110$\pm$0.006  & 0.417$\pm$0.014   & 0.132$\pm$0.009    \\
0.15-0.20 & 0.400$\pm$0.011 & 0.103$\pm$0.005  & 0.393$\pm$0.014   & 0.117$\pm$0.006    \\
0.20-0.25 & 0.349$\pm$0.011 & 0.087$\pm$0.005  & 0.361$\pm$0.016   & 0.119$\pm$0.005    \\
0.25-0.30 & 0.313$\pm$0.011 & 0.093$\pm$0.006  & 0.348$\pm$0.017   & 0.125$\pm$0.006    \\
0.30-0.40 & 0.287$\pm$0.010 & 0.095$\pm$0.005  & 0.322$\pm$0.014   & 0.139$\pm$0.006    \\
0.40-0.50 & 0.261$\pm$0.020 & 0.120$\pm$0.007  & 0.285$\pm$0.017   & 0.132$\pm$0.007    \\
0.50-0.70 & 0.252$\pm$0.022 & 0.122$\pm$0.010  & 0.266$\pm$0.021   & 0.158$\pm$0.007    \\
0.70-1.00 & 0.278$\pm$0.053 & 0.093$\pm$0.023  & 0.286$\pm$0.057   & 0.275$\pm$0.035    \\
\hline
\end{tabular}
\end{table}

\subsection{Abundance profiles}
The projected temperature and metallicity profiles are plotted in Fig. 4, scaled by $R_{500}$. The temperature profiles ({\it top-left panel}), renormalized by the average temperature estimated within 0.3$R_{500}$,  behave quite universally increasing from the center and reaching the maximum at 0.1-0.2$R_{500}$ and a slow decline beyond that peak. The metallicity profiles, show a large scatter in the center with values ranging between 0.3 and 2 Z$_{\odot}$,  but show also a universal decrease with radius with a flattening beyond $\sim$0.4$R_{500}$.  
In Fig. 5 we show the stacked profiles that have been estimated using Monte Carlo simulations. We performed 10,000 realizations of the profiles by randomly varying  the observational data points of the temperature and metallicity profiles to determine new distributions. The randomization was derived from the distribution of measurement values  and errors. The randomization of the data points with the radius, scaled by $R_{500}$, was done using a truncated gaussian distribution to bound the points to the extraction area. For each realization and each radial bin we computed the average metallicity and the scatter to obtain a distribution of values. The median values of 10,000 realizations are shown in Fig. 5 with the 68$\%$ uncertainties taken from the distributions. In Appendix D we compare the stacked profile derived here with the one derived using the weighted mean.
The stacked metallicity profile ({\it right panel} of Fig. 5) is compared with what was found by \cite{2008A&A...487..461L} who analyzed a sample of massive systems (i.e. kT$>$3 keV) and by \cite{2017A&A...603A..80M} who analyzed a sample of cool-core systems mostly in the low-mass regime. Our average profiles are in quite good agreement with the results of \cite{2008A&A...487..461L}. The slightly higher metallicity obtained in their work can be easily explained by the use of a different N$_H$ (\citealt{2008A&A...487..461L} used the LAB values instead of the total N$_H$ used in this work). A higher N$_H$ (as assumed in this work) returns in general a smaller metallicity value, mainly due to the change in the measured temperature. For average column densities (3-5$\times10^{20}$ cm$^{-2}$) the impact can be of the order of 5-10$\%$, enough to compensate for the observed difference. See Appendix A for more details. 

If we exclude the inner and outermost data points our metallicities are systematically lower (5-15$\%$) than the ones by \cite{2017A&A...603A..80M}. This deviation do not arise from a different column density because  the total N$_H$ values were also used by \cite{2017A&A...603A..80M} when fitting the data. Indeed their sample by construction includes only relaxed, cool-core systems while our sample includes also very disturbed systems. To understand if these disturbed systems can justify the observed difference, we split the sample based on their dynamical state. To emphasize the difference we only consider relaxed  (i.e. the ones in the bottom-right quadrant of Fig. 1) and disturbed (i.e. the ones in the top-left quadrant) clusters.  The result is shown in Fig. 6 and summarized in Table 1. Indeed, the relaxed systems showed a strong drop from the center to $\sim$0.2-0.3 $R_{500}$ where the profiles flatten. On the contrary, disturbed systems show a much shallower drop. However, even when comparing only the relaxed systems there is still some tension between our results and the average profiles obtained by \cite{2017A&A...603A..80M} beyond the central bin. Thus, the difference  cannot entirely be explained by the presence of a larger fraction of disturbed systems in our sample. Differences may also arise from the use of different spectral fitting packages and plasma codes (e.g. \citealt{2017A&A...603A..80M} performed their analysis using the SPEX package, see \citealt{1996uxsa.conf..411K}, and un updated version of the MEKAL code, see \citealt{1985A&AS...62..197M}). So, a more detailed investigation is required to understand the cause of this difference but it is beyond the scope of this paper.

The scatter decreases almost monotonically from the center to the outer regions for relaxed systems while for disturbed clusters it reaches a minimum at $\sim$0.2-0.3$R_{500}$ and then it increases again in the outer regions (where anyway we have only a few measurements). The scatter in the cores is much larger for relaxed systems while disturbed systems show a larger scatter beyond $\sim$0.2$R_{500}$. The increase of the scatter in the outer regions can be seen in Fig. 7 where we showed the metallicity profiles normalized by $A_{core}$. Relaxed clusters (blue points) behave very similarly, with the highest value in the center and a drop of already a factor of 2 at $\sim$0.1$R_{500}$ for most of the systems. Beyond $\sim$0.4$R_{500}$ the metallicity of relaxed systems is a factor of 4-5 smaller than in the center. For disturbed systems instead is not always a monotonic decrease of the metallicity.  The drop from the center is not uniform from cluster to cluster with the different profiles showing a very inhomogeneous distribution with the effect to significantly increase the  scatter  in the outer regions. In some cases the outer metallicity values are higher than the values measured in the center. This is the case for $\sim$30$\%$ of the clusters in the top-left quadrant (disturbed) and for a few in the top-right (peaked-clumpi) or bottom-left (smooth) quadrants.  The only cluster that was classified as relaxed and shows an increasing metallicity in the outer regions  is A2142, a massive cluster that shows a significant  dynamical activity at large scales as shown by, e.g.,  \cite{2011ApJ...741..122O}, \cite{2013A&A...556A..44R}, and \cite{2016A&A...595A..42T}.

\begin{figure}
\includegraphics[width=1.1\columnwidth]{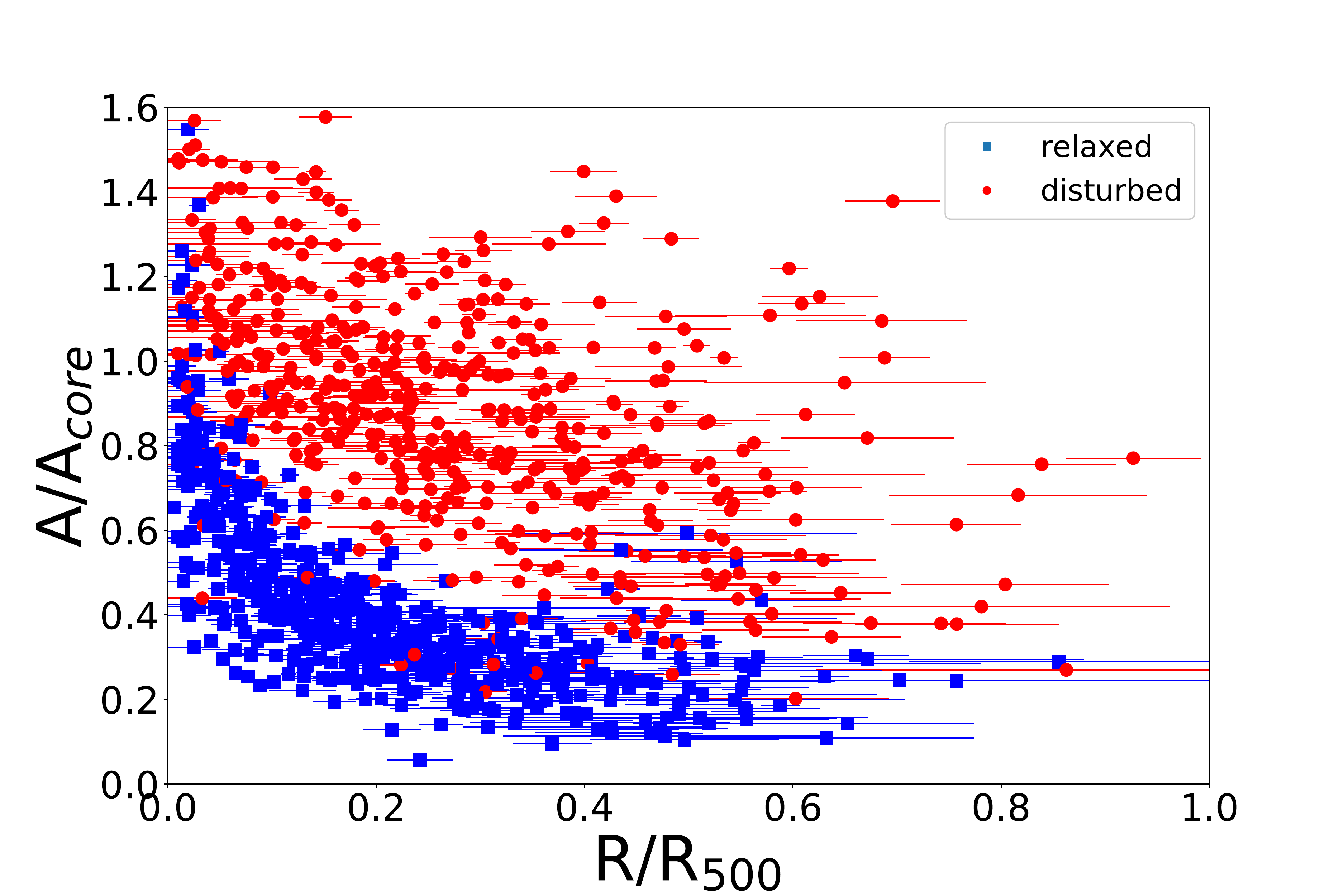}
\caption{Metallicity profiles normalized by $A_{core}$. Relaxed systems are shown in blue while disturbed clusters are shown in red. Errorbars in the Y axis are not shown.}
\end{figure}

\subsection{AGN feedback}
AGNs have been shown to be an efficient mechanism for uplifting the metals from the central regions to the outskirts of a few individual galaxy clusters (e.g. \citealt{2009A&A...493..409S}, \citealt{2012ApJ...753...47D}, \citealt{2015MNRAS.452.4361K}). Using hydrodynamical simulations, \cite{2017MNRAS.468..531B} showed that AGNs are indeed expected to distribute the metal-rich gas out to large radii. Here, with a large sample of galaxy clusters we statistically investigate the impact of AGN feedback on the metallicity profiles subdividing our sample in clusters with and without a central radio source (CRS). The presence of the CRS was identified using the NVSS catalog by requiring a maximum separation of 50 kpc from the X-ray peak as suggested by \cite{2009A&A...501..835M} (see also \citealt{2007MNRAS.379..100E}). Both distribution of clusters with and without a CRS cover a large range of temperatures (see Fig. 2). 
For the comparison of the metallicity profiles we only used the relaxed systems of our sample. This ensures that any effect is not related to the different profiles of relaxed and disturbed systems shown in the previous section. We also verified that even for this subsample, both galaxy clusters with and without a CRS span a broad temperature range ensuring that any effect is independent of the weak dependence of the metallicity with the temperature discussed in Section 3.1.  The results are shown in Fig. 8 and summarized in Table 2.   The profiles of the systems hosting a CRS are much flatter, i.e. in the center the metallicity is systematically lower, at 10-20$\%$ level (although the scatter is also large), than the ones without a CRS. The effect is expected to be much stronger at the groups scale because AGNs are expected to leave stronger imprints at the galaxy group scale due to their shallower potential. Unfortunately, although there is a hint for a steeper profiles for the groups without a central radio source we only have 5 systems and that  question should be confirmed with a larger sample.

\begin{table}
\renewcommand{\thetable}{\arabic{table}}
\centering
\caption{Average metallicity profiles for galaxy groups and clusters with or without central radio source.}
\begin{tabular}{|c|ccc|ccc|}
\hline
\multicolumn{1}{|c|}{} & \multicolumn{2}{c|}{RS} & \multicolumn{2}{c|}{no RS}  \\
\hline
Radius & A   & scatter  & A  & scatter \\
(r/r$_{500}$) & (solar) & &  (solar)  &\\
\hline
0.00-0.05 & 0.782$\pm$0.015 & 0.215$\pm$0.010  & 0.900$\pm$0.064   & 0.359$\pm$0.018    \\
0.05-0.10 & 0.560$\pm$0.012 & 0.141$\pm$0.008  & 0.654$\pm$0.032   & 0.157$\pm$0.023    \\
0.10-0.15 & 0.446$\pm$0.010 & 0.093$\pm$0.005  & 0.476$\pm$0.036   & 0.154$\pm$0.017    \\
0.15-0.20 & 0.397$\pm$0.011 & 0.089$\pm$0.005  & 0.386$\pm$0.039   & 0.120$\pm$0.013    \\
0.20-0.25 & 0.349$\pm$0.012 & 0.076$\pm$0.004  & 0.334$\pm$0.026   & 0.086$\pm$0.008    \\
0.25-0.30 & 0.315$\pm$0.012 & 0.083$\pm$0.006  & 0.320$\pm$0.029   & 0.106$\pm$0.009    \\
0.30-0.40 & 0.288$\pm$0.010 & 0.087$\pm$0.004  & 0.292$\pm$0.023   & 0.098$\pm$0.009    \\
0.40-0.50 & 0.262$\pm$0.022 & 0.118$\pm$0.007  & 0.261$\pm$0.041   & 0.103$\pm$0.017    \\
0.50-0.70 & 0.252$\pm$0.025 & 0.118$\pm$0.009  & 0.225$\pm$0.071   & 0.150$\pm$0.059   \\
0.70-1.00 & 0.293$\pm$0.029 & 0.091$\pm$0.025  & -   & -   \\

\hline
\end{tabular}
\end{table}

\begin{figure}
\includegraphics[width=0.75\columnwidth,angle=270]{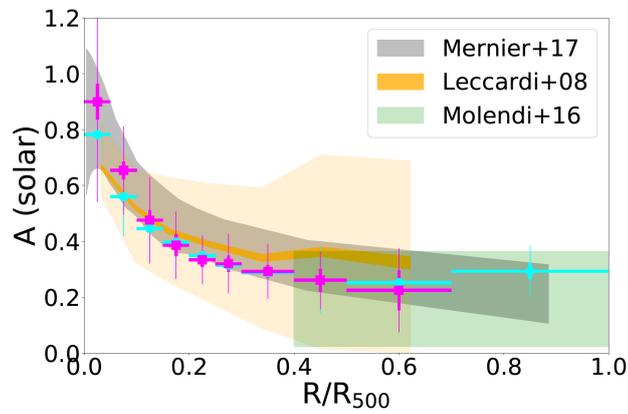}
\caption{Stacked average metallicity profiles for galaxy groups and clusters with (cyan points) or without (magenta points) a central radio source. We used only the relaxed systems (i.e. the bottom-right panel of Fig. 1). }
\end{figure}

\begin{figure}
\includegraphics[width=1.05\columnwidth]{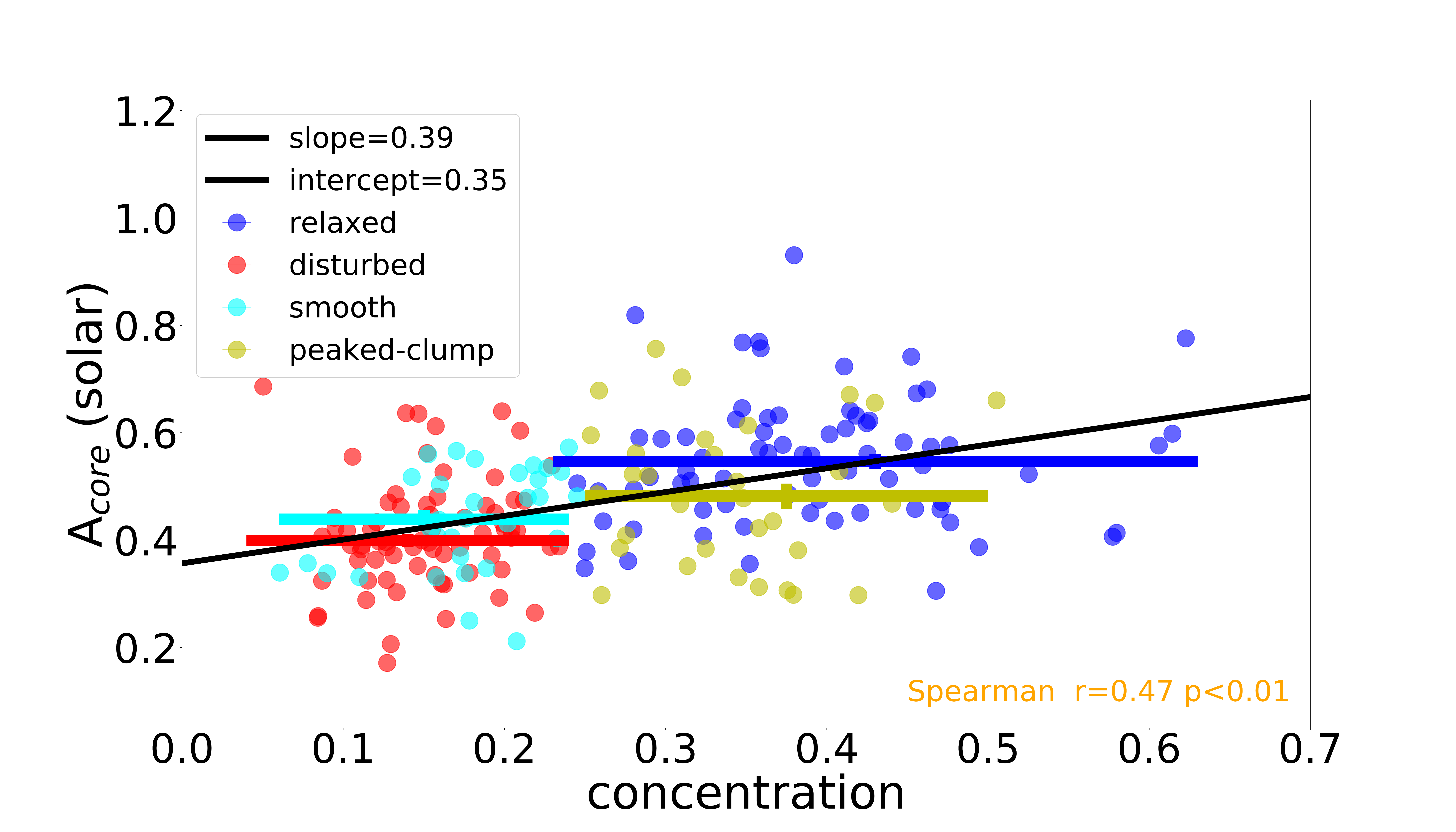}
\vspace{10pt}
\includegraphics[width=1.05\columnwidth]{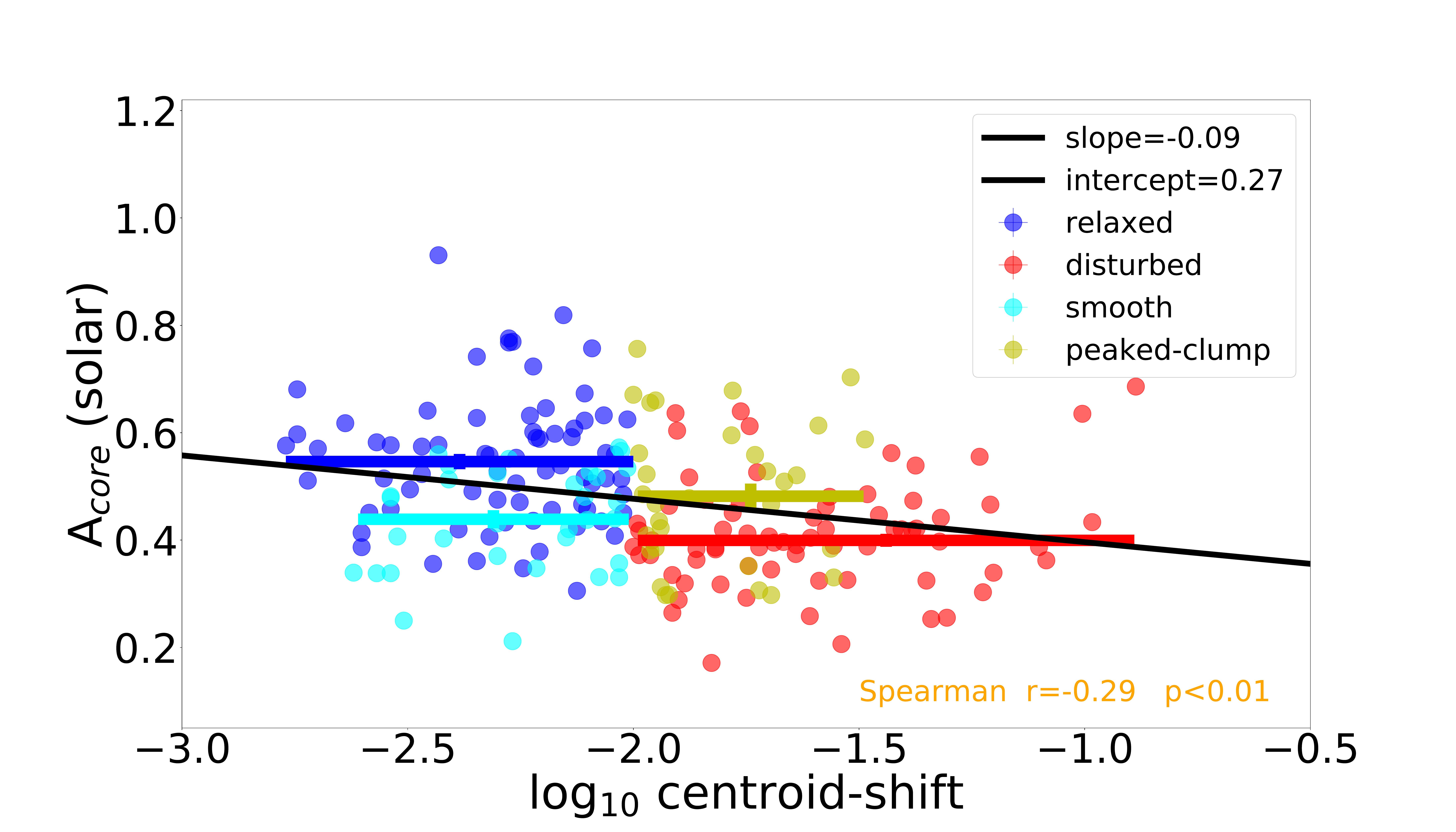}
\includegraphics[width=\columnwidth]{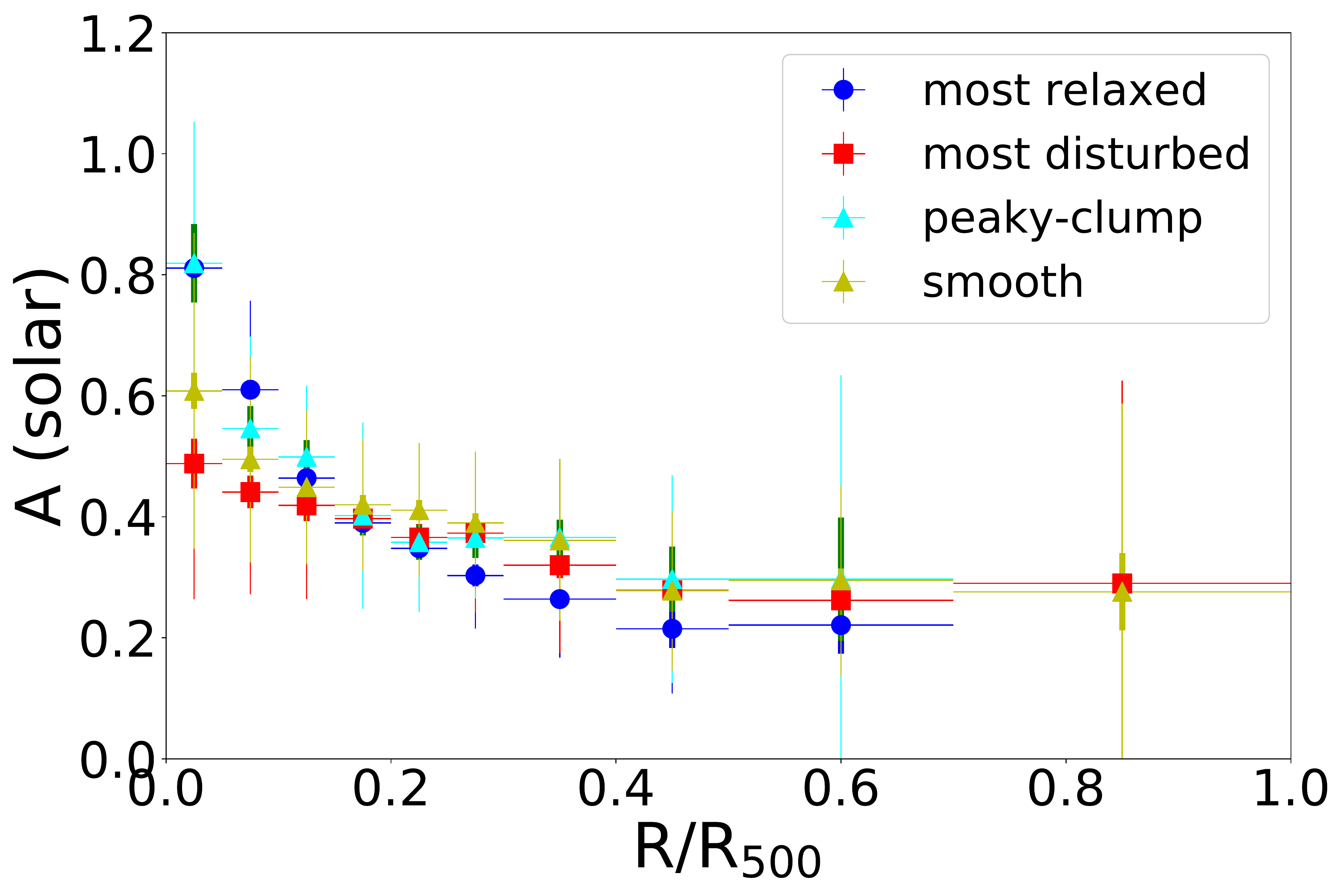}
\caption{{\it Top:} abundance-concentration plot. The data points are the individual measurements and the colors are the same as in Fig. 3 with the bold lines representing the median of the distribution for each subsample. The core abundances have been derived within 0.3$R_{500}$ and the concentration within 0.5$R_{500}$. The cyan line is the best fit to the individual data points. The measurement errors for both the concentration and the core abundance are not shown for visualization purposes but they are accounted for in the fit. {\it Middle:} the same as in the top-panel but for the centroid-shift. {\it Bottom:} stacked metallicity profiles for groups and clusters populating a different region of the concentration-centroid shift diagram as shown in the inset plot. The most relaxed clusters are the ones with a concentration above the 67 percentile and a centroid-shift lower than the 33 percentile. The most disturbed clusters are the ones with a concentration below the 33 percentile and a centroid-shift higher than the 67 percentile. Peaky-clumpi and smooth clusters are as defined in Sect. 2.5.
 }
\label{fig:evolving}
\end{figure}

\subsection{Shaping the abundance profile in the cores}
The shape of the metallicity profiles depends on the individual cluster histories. However, if we use the concentration-centroid shift diagram as a proxy of the state of relaxation of a cluster, we can study how the central abundance and shape of the profiles change as the dynamical state varies. 

Despite the large scatter, morphological parameters are a measure of how relaxed the cluster is. Thus, for example the higher is the concentration the more relaxed is expected to be the cluster and the more time is probably passed since the last major merger. As a consequence, the expected abundance in the core of these clusters is expected to be higher because it hasn't been mixed and redistributed by large scale motions induced through the merger. In Fig. 9 ({\it top panel}) we show the positive correlation  (slope=0.39$\pm$0.06, scatter=0.109$\pm$0.04) between the abundance and the concentration. The moderate correlation is confirmed also by the Spearman rank test (r=0.47, p$<$0.01). The most peaked clusters have on average a higher metallicity in the center. The tendency of having the highest metallicity in the most relaxed systems is also confirmed by the negative correlation between the abundance and the centroid-shift (middle panel).

While some clusters are clearly relaxed (e.g. high concentration and low centroid-shift) or disturbed (low concentration and high centroid-shift) there are some clusters that are not so easily defined. As discussed in Sect. 2.5 they can be interpreted as clusters evolving in one or another dynamical state (i.e. relaxed objects which show infalling substructures, or post merger objects that are slowly moving back to the relaxed status).  In the bottom panel of Fig. 9 we compare the metallicity profiles of the most relaxed (clusters with a concentration above the 67 percentile and a centroid-shift below the 33 percentile) and most disturbed objects  (clusters with a concentration below the 33 percentile and a centroid-shift above the 67 percentile) with the peaky-clumpi and smooth systems. Since there are a few low temperature  systems with very high metallicity (see Fig. 2) to make the plot we excluded all the systems with kT$<$2 keV to make sure that any observed trend is not associated with them. Apart from the most relaxed systems which show a hint of lower metallicity, beyond $\sim$0.3$R_{500}$ all the profiles are basically flat and behave very similarly. However, in the core they look quite different, with the metallicity profiles steepening moving from the most disturbed to the most relaxed.  This behavior is confirmed also when cutting at higher temperatures. Moreover, we note that while relaxed and peaky clusters share a similar core abundance they have a different metallicity profile. In the innermost bin the metallicity is very similar (both still have a core) but the most relaxed systems have a much steeper profile.

\section{Discussion}
\subsection{Whole core abundances}
Our analysis of the global abundances is based on the inner regions of galaxy groups and clusters. We find that this choice emphasizes the differences between systems with different properties (e.g. temperature, dynamical state). In fact the largest difference in the metallicity profiles arises  in the innermost regions while in the outskirts the divergence is not significant (in part because the quality of the data and the limitation of the current instruments do not allow to measure eventual small metal variation).  By fitting the data points of the individual measurements we obtained the weak dependence of the metallicity with cluster mass in agreement with what found by previous studies (e.g., see  \citealt{2007A&A...462..429B}; \citealt{2017MNRAS.464.3169Y}).  The higher metallicity  in low-mass systems together with their lower gas mass with respect of hot clusters (e.g. \citealt{2015A&A...573A.118L}) was interpreted as an indication for a lower star-formation efficiency in massive systems (e.g., \citealt{2011A&A...535A..78Z}). However, this metal dependence disappears when only the disturbed systems are considered. There is still a mild dependence for relaxed systems but it is possibly dominated by a very few low-temperature objects with extremely high metallicity. In fact, when removing low kT systems the dependence is much more uncertain. This agrees well with the result by \cite{2018MNRAS.tmpL..84M} who found a very similar average Fe enrichment within 0.1$R_{500}$ for a sample of 44 relaxed elliptical galaxies, groups, and clusters. Thus, a different fraction of relaxed and disturbed clusters at different temperatures in the analyzed samples might lead to a different observed dependence. Anyhow, we note that while there are 32 groups and clusters with a global metallicity within 0.3$R_{500}$ higher than 0.6 solar for temperatures below 4 keV,  there is only one cluster (out of 13 relaxed systems) with such a high abundance at high temperatures.  One possible explanation is that it is very difficult to build a very massive galaxy cluster without undergoing a major merger (or multiple minor mergers). During these events the core is disrupted or strongly affected, with the final effect of lowering the central metallicity. N-body simulations find that small clusters and groups have a lower merger rate (e.g., \citealt{2008MNRAS.388.1792N},  \citealt{2009ApJ...701.2002G}),  so that their central metallicity patterns  are not destroyed during their evolution, and they can potentially reach high values. Another possible explanation is that in massive galaxy clusters the contribution to the enrichment from galaxies is probably negligible because galaxies comprise less of $5\%$ of the baryons in
rich clusters. At the groups scale instead the baryon fraction in galaxy members is at least equal to that of the hot gas (\citealt{2009ApJ...703..982G}), so enriched material from galaxies can significantly impact the amount of metals present in the ICM. That would explain the slightly higher metallicity observed in the center of the low-mass systems. However, low-mass systems also show in some cases a quite low metallicity (Z$\le$0.3Z$_{\odot}$) which is not observed at high temperatures. While there are possible systematic uncertainties related to multitemperature structure, this might indicate that indeed different processes are at work in different systems and that for some reason the star-formation efficiency in some groups is not efficient. These low-metallicity clusters are almost equally split in clusters with and without a CRS so the presence/absence  of central AGN is probably not the cause of the high scatter observed in the low-mass regime. 

\subsection{Average profiles}
Fig. 4 shows the individual projected temperature ({\it left panel}) and metallicity ({\it right-panel}) profiles as a function of the radius rescaled by $R_{500}$.  It can clearly be seen that the dispersion is much larger in the core where some individual systems show a strongly peaked metallicity while others show lower values. This large scatter is independent of their dynamical state. The metallicity then declines with radius and seems pretty constant (within the large scatter) beyond 0.3-0.4$R_{500}$. Fig. 5 shows the average profiles, split by their dynamical state. 
While the metallicity profiles of relaxed and disturbed systems differ in the central regions (R$<$0.1$R_{500}$), at large radii (i.e. beyond 0.4$R_{500}$) they are consistent with a flat distribution: a fit with a constant for R$>$R$_{500}$ gives a value of 0.28$\pm$0.11 $Z_{\odot}$ (but see \citealt{2016A&A...586A..32M} for a description of the challenges in determining the metallicity in the outskirts). 
In the 0.3-0.4$R_{500}$ region the average profile of disturbed systems shows hints of a metallicity excess with respect to the average profile of relaxed clusters. This effect was also obtained with hydrodynamical simulations by \cite{2017MNRAS.468..531B}. It is possibly due to the metals that have been spread out from the center during the merging. The difference of the core metallicity as function of the cluster dynamical state, and the convergence of the profiles for relaxed and disturbed systems at large radii was also observed by \cite{2017MNRAS.472.2877M}.

The difference between the profiles of relaxed and disturbed systems was already observed by \cite{2001ApJ...551..153D} who analyzed a sample of 17 hot systems with ASCA data. Their results showed a much larger metallicity gradient for cool-core systems than non-cool-core clusters.  The moderate gradient observed for disturbed systems was interpreted as the remnant of a much stronger gradient that has not been completely erased by the merger events. Our results support this scenario as shown in the bottom panel of Fig. 9.  The most relaxed objects share the same metallicity in the center of the peaky-clumpi objects in agreement with the fact that the cores of both subsamples are unaffected. However the metallicity profile for the  most relaxed clusters  is much steeper than the one of the peaky-clumpi clusters for which the disturbance detected with the morphological parameters is probably associated with a large scale mixing.  The most disturbed objects (in red) show a very weak gradient of the metallicity in the center, while the smooth systems  (in green) show hint of a    metallicity increase toward the center. Thus, there is a strict connection between the observed metallicity profiles and the morphological properties of the systems. Major mergers are able to erase the metallicity pattern in the center but also minor mergers have an impact on the metallicity profiles. However, even in presence of a major merger we might still be seeing a moderate metallicity peak in the center if we are in the very early merger stage where the metals have not been mixed yet, or in an old merger stage where a new abundance gradient is forming again. 

\subsection{AGNs feedback}
The feedback from the central supermassive black holes has an impact on the distribution of the hot ICM, so they can potentially affect the shape of the metallicity profiles. AGNs manifest as central radio sources, which jets are often responsible for disturbances of the ICM. One example are the so-called X-ray cavities associated with inflating lobes of radio-emitting plasma. Some of the work done by these cavities is used to lift the metal-enriched gas (e.g. \citealt{2015MNRAS.452.4361K}) implying that AGNs may be able to mix the metals on scales of several hundreds of kpc.  In Fig. 8 we compared the metallicity profile for relaxed systems hosting a central AGN with the one without a central AGN and we found that indeed the former has a much flatter profile.  This is interpreted as the redistribution of the enriched gas due to the AGNs. Another hint to support this hypothesis is the different impact of the AGNs on the profiles of galaxy groups and galaxy clusters, with the latter less affected by the presence of the CRS in the center. The effects of feedback from central AGNs is in fact expected to be more important at the scale of groups because of their shallower potential wells. It would be important to confirm this apparent trend with an even larger sample.  
We observe hints that profiles of groups and clusters with a central AGN have also a higher metallicity in the 0.2-0.4$R_{500}$ region.  This is the first observational confirmation, albeit still at low significance, for the prediction of hydrodynamical simulations by \cite{2017MNRAS.468..531B}, who found that  including the AGN feedback in the hydrodynamical simulations strongly reduce the central metallicity and increases the metallicity in the outer regions.

\subsection{Low- and high-mass systems}
\cite{2016A&A...592A.157M,2017A&A...603A..80M} reported a much lower metallicity outside the core in galaxy groups with respect to the one in galaxy clusters while in the core they found a similar abundance. The median temperature of their sample was 1.7 keV and it is the value they used to characterize groups and clusters.  Since we excluded all the nearby systems for which the XMM-Newton FOV is not large enough to measure at least 0.3$R_{500}$, our sample does not include most of the galaxy groups analyzed by \cite{2016A&A...592A.157M,2017A&A...603A..80M}. Moreover, by construction we did not analyze bright and nearby elliptical galaxies, so the number of systems in our sample with an average temperature lower than 1.7 keV is limited to only 18 galaxy groups. A more reasonable value for our sample is a cut at 3 keV, so that the groups and cluster subsamples have roughly a similar size. Moreover, since in the previous section we found that the presence/absence of a CRS may impact the metallicity profiles, for each subsample (i.e. groups and clusters) we investigate objects with and without a CRS separately. Despite the large scatter, in both cases we found that galaxy groups have a higher metallicity than galaxy clusters within $\sim$0.1$R_{500}$.  The metallicity in groups drops quite fast and galaxy clusters show a higher metallicity beyond that radius in agreement with the finding by  \cite{2016A&A...592A.157M,2017A&A...603A..80M}. The higher metallicity in the outer regions of galaxy clusters is also confirmed if we only consider the relaxed systems (i.e. bottom-left quadrant in Fig. 1) to avoid any bias from the metallicity-dynamical state connection discussed in Sect 4.2.

\section{Conclusions}
We derived the average metallicity within 0.3$R_{500}$ for 207 groups and clusters in the local Universe ($z$$<$0.1) observed with XMM-Newton. For 156 systems the data were sufficient also to determine the averaged azimuthal metallicity profiles up to a minimum of 0.4$R_{500}$ and a maximum of $\approx$$R_{500}$.  
Our main findings are the following:
\begin{itemize}
\item[$\bullet$] A mild anti-correlation between the average metallicity and temperature within 0.3$R_{500}$.  However, relaxed systems have typically higher mean metallicity than disturbed objects and when fitting the abundance-temperature correlation for relaxed and disturbed objects independently this mild anti-correlation weakens/disappears. 
\item[$\bullet$] The metallicity profiles rapidly decrease from the center to 0.2-0.3$R_{500}$ where they flatten and stay constant out to large radii. While behaving very similarly beyond 0.1-0.2$R_{500}$  the profiles for relaxed and disturbed systems diverge in the center where the formers have a much more peaked distribution. 
\item[$\bullet$] The average profile determined for relaxed systems hosting a CRS is flatter than the one determined for relaxed systems without a CRS. We interpret this in the sense that central AGNs can modify the shape of the metallicity profiles. Supportive of this interpretation is that  the difference between the profiles with and without CRS is more accentuated at the galaxy groups scale where AGNs feedback is thought to be stronger due to their lower potential wells. 
\item[$\bullet$] Using the concentration-centroid shift diagram to classify different stages of the clusters' dynamical state we found that the more relaxed the clusters are, the more the metallicity profiles become steeper in the cores. We argued that this depends on how much time has passed since the last major merger: shortly after the merging the metallicity pattern is completely erased but if nothing new happens than the metallicity profile slowly begins to rebuild. 
\item[$\bullet$] Galaxy groups and clusters with similar properties (e.g. with or without CRS) have a different metallicity profiles suggesting that the mechanisms at work to enrich and mix the hot gas may not be the same at all mass scales.
\item[$\bullet$]  We showed that any trend depends on the applied sample selection and that, consequently, the sample needs to be well-characterized (in terms of mass, morphological state, central AGN) to enable meaningful interpretations.
\end{itemize}

\section*{Acknowledgements}
We thank the anonymous referee for the useful report which helped to improve the quality of the paper. We acknowledge Gerrit Schellenberger for the help and fruitful discussion. 
LL and THR acknowledge support from the German Research Association (DFG) through the Transregional Collaborative Research Centre TRR33 ``The Dark Universe'' (project B18). LL also acknowledges support from NASA through contract NNX17AD83G.




\bibliographystyle{mnras}
\bibliography{metallicity} 




\appendix
\section{Impact of the used column density on the measured abundance}
\begin{figure}
\includegraphics[width=0.5\textwidth]{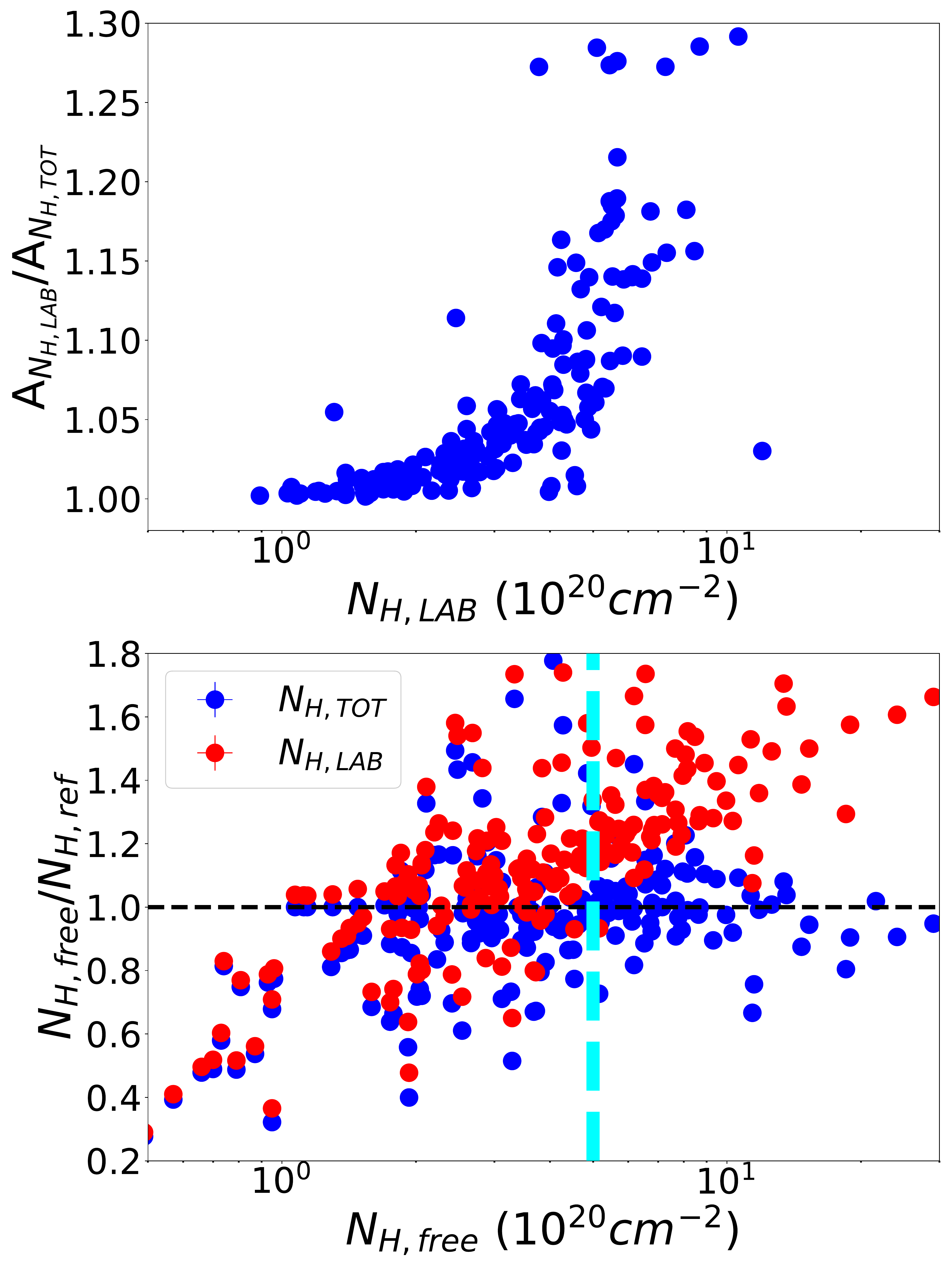}
\caption{{\it  top:}  the ratio between the abundance derived using $N_{\rm H,LAB}$ and $N_{\rm H,TOT}$ as function of the LAB value.  ({\it bottom})  ratio between the measured $N_{\rm H}$ value (i.e. left free to vary in the fit) and $N_{\rm H,LAB}$ (red points) and $N_{\rm H,TOT}$ (blue points), respectively. The agreement between the fitted value and $N_{\rm H,TOT}$ is quite good, in particular above $\sim$5$\times$10$^{20}$ ${\rm cm^{-2}}$ (cyan line) when fitting  $N_{\rm H}$ becomes less challenging.  }
\end{figure}
X-rays can be absorbed by material along the line of sight. It follows that accurate cluster property measurements requires a careful estimation of the absorption. Commonly this effect is quoted as the equivalent column density of hydrogen although generally it is due to heavier elements. The abundance of these metals is usually traced by measuring the neutral hydrogen HI column density with  21cm measurements. A frequently used  HI-survey is the Leiden/Argentine/Bonn (LAB) survey (\citealt{2005A&A...440..775K}). This only provides the neutral hydrogen contribution along the line of sight but does not account for the molecular and ionized hydrogen. A measure of the molecular hydrogen has been provided by \cite{2013MNRAS.431..394W} using the dust extinction E(B-V) measured in the B and V band and calibrated using the X-ray afterglows of Gamma Ray Bursts. 
As shown by, e.g., \cite{2015A&A...575A..30S} the contribution of the molecular component starts to be significant above $\sim$5$\times10^{20}$ cm$^{-2}$ and can have a significant impact on the measured temperatures, and therefore, also on the abundances. In Fig. A1 ({\it top panel}) we show how the measured metallicity varies by using the LAB or total N$_H$. While for low N$_H$ values (i.e. $\le3\times10^{20}$ cm$^{-2}$) the effect is almost negligible, for higher values (i.e. 3-6$\times10^{20}$ cm$^{-2}$)  can be as high as 10$\%$. For higher column densities the impact is even more dramatic and can reach $50\%$ or more. 

In Fig. A1 ({\it bottom panel}) we compare the column densities obtained by leaving $N_{\rm H}$ free to vary during the fit of the spectra extracted within $R_{500}$, with $N_{\rm H,LAB}$ and $N_{\rm H, TOT}$, respectively.  Indeed, the agreement between the fitted values and $N_{\rm H, TOT}$ is much better, in particular for large $N_{\rm H}$ values. We note that in a few cases, even using $N_{\rm H, TOT}$ is not enough and leaves clear residuals in the spectral fit. In these cases we use the $N_{\rm H, FREE}$ to determine the spectral properties used in this paper. At low column densities the fitted $N_{\rm H}$ values are in general lower (in agreement with the finding by \citealt{2015A&A...575A..30S}) than the tabulated $N_{\rm H}$ but the uncertainties in the measurements are pretty large.

\section{S/N for the metal profiles}
\begin{figure*}
\includegraphics[width=\textwidth]{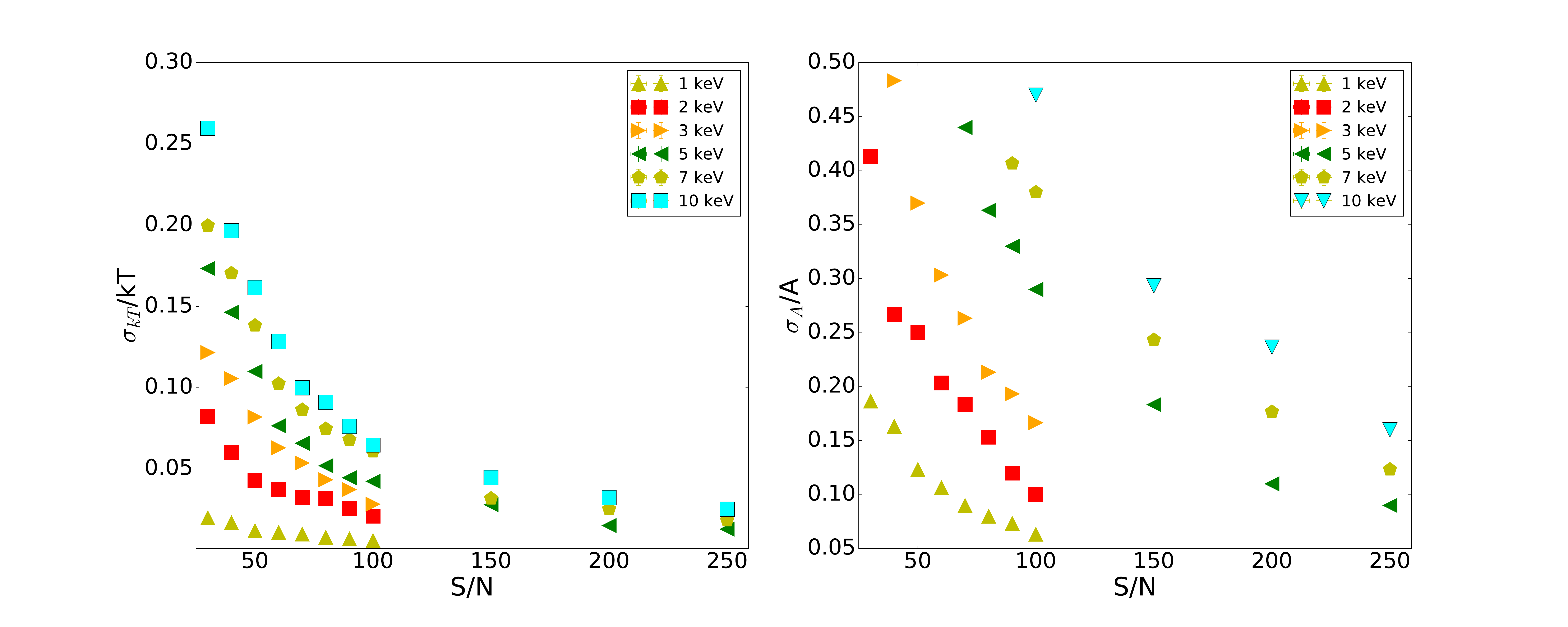}
\caption{{\it  left:} The relative error of the temperature ({\it left}) and metal ({\it right}) measurements as a function of the S/N for clusters with different temperatures.  These are just statistical errors. For plotting purposes we only show the results for 5 input temperatures. \label{fig:ratio}}
\end{figure*}
Due to the fading of the emission lines, it is harder to estimate the metallicity of the high temperature galaxy clusters than the one of cooler galaxy groups. To have a fairly similar selection of the annuli for galaxy clusters with different temperatures we ran a set of simulations to estimate the required S/N necessary to have an estimate of the metallicity better than $\sim$30$\%$. 

We assumed an extraction region of 5 arcmin, an exposure time of 30 ks,  redshift=0.05, N$_{H}$=3$\times10^{20} {\rm cm}^{-2}$, and  an input metallicity of 0.3 solar. 
The model of the particle background components has been determined using the FWC observations while the foreground emission using the results from \cite{2002A&A...389...93L} and rescaled to an area of 5$^{\prime}$ radius.
Then, the normalization of the cluster thermal component was modified to match the required S/N. Each combination was ran 1000 times and for every realization we computed the relative error for the temperature and metallicity. In Fig. B1 we show the median values of the distribution of relative errors with the 68$\%$ errors taken from the distributions. For cool systems (e.g. kT$<$3 keV) a S/N of 50 is sufficient to get metallicity measurements better than 30$\%$. For hotter systems we had to use a S/N of 100 or 150. 

Note that for most of the clusters in our sample the S/N in the innermost annuli is still much higher than our requirement because of the minimum bin size of 30 arcsec. We checked our measured values to make sure that the fitted relative errors are consistent with the expectation from our simulations, and indeed the agreement is fairly good although in the outermost bins we usually find slightly larger relative errors. This is possibly due to the stronger fluorescence lines in the outer CCDs compared to the ones determined in the central CCD and that were assumed for our simulations. Anyway, that does not affect in any way the conclusion of the paper.

\section{Spectral fitting}
As explained in Sect. 2.4, for each cluster we are doing a joint fit of our XMM-Newton spectra and the RASS spectrum obtained with the X-ray background tool by HEASARC, for which we cannot use the cstat statistic\footnote{In a recent release the support for counts statistics and the creation of a counts-based spectrum has been added to the tool.} in Xspec. Several studies discussed that the use of the $\chi^2$ statistic may bias the results (e.g. \citealt{2009ApJ...693..822H}), although the effect on the metallicity seems smaller than the one on the temperature for a large range of S/N (e.g. \citealt{2017MNRAS.472.2877M}). To investigate whether our results suffer from a measurement bias we ran a set of simulations. We simulated spectra using an exposure of 20 ks for $kT$=1, 3, and 5 keV, a metallicity of 0.5, and a redshift of 0.05. The model for the background has been added following the procedure described in Appendix B. Then, the normalization of the cluster component was modified to match the interesting S/N: for 1 keV plasma we investigated spectra with a S/N=30, 50, and 100 while for hotter systems S/N=50, 100, and 150.  Each combination was ran 1000 times. In Fig. D1 we show the median of the best fit values obtained by fitting the spectra with cstat or $\chi^2$ statistic. When using $\chi^2$ we fitted the spectra with standard binning (i.e. minimum of 25 counts per bin, referred as group in the figure) and spectra binned to get a minimum energy width of each group of at least 1/3 of the full width half maximum (FWHM) resolution at the central photon energy of the group (refereed as specgroup in the figure). The second choice strongly decreases the number of bin that will be fitted. 
We find that biases in the abundance determination in the S/N regime of interest (see appendix B for the choice of the S/N) using the $\chi^2$ statistic are in agreement with what obtained using cstat, in particular when the grouping is done by requiring a minimum energy resolution. This is because the bias is expected to increase with the ratio between the square root of the number of counts and the number of bin (i.e. $\sqrt{N_C}/N_{bin}$, see \citealt{2009ApJ...693..822H}). While with standard grouping, $N_{bin}$ always increases with increasing $N_C$ this is not the case when setting a minimum energy width for the binning. Thus, when the number of counts is high (which is almost always the case in our spectra thanks to the high S/N required) the $\sqrt{N_C}/N_{bin}$ ratio becomes larger than 1 and the bias is comparable with what one obtain with cstat.

\begin{figure}
\includegraphics[width=\columnwidth]{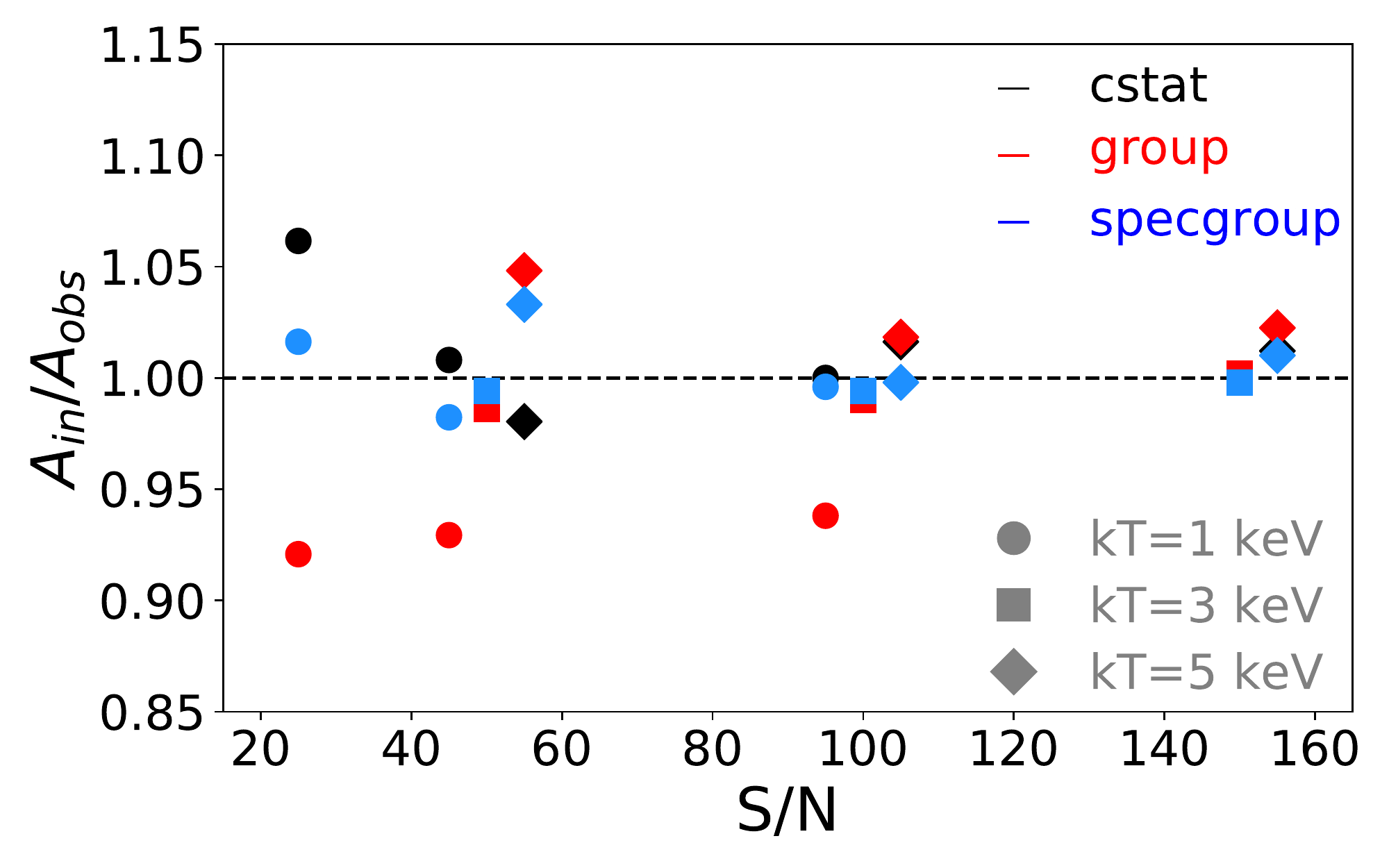}
\caption{The points are the median of the best fit values of the metallicity divided by the input metallicity obtained for each combination of temperature, S/N, and fitting method. Each combination was ran 1000 times.  The circles refer to kT=1 keV, the squares to kT=3 keV, and diamonds to 5 keV plasma. Circles and diamonds have been shifted by -5 and +5 in S/N for visualization purposes}
\end{figure}

\section{Weighted mean vs median metallicities}
\begin{figure}
\includegraphics[width=0.5\textwidth]{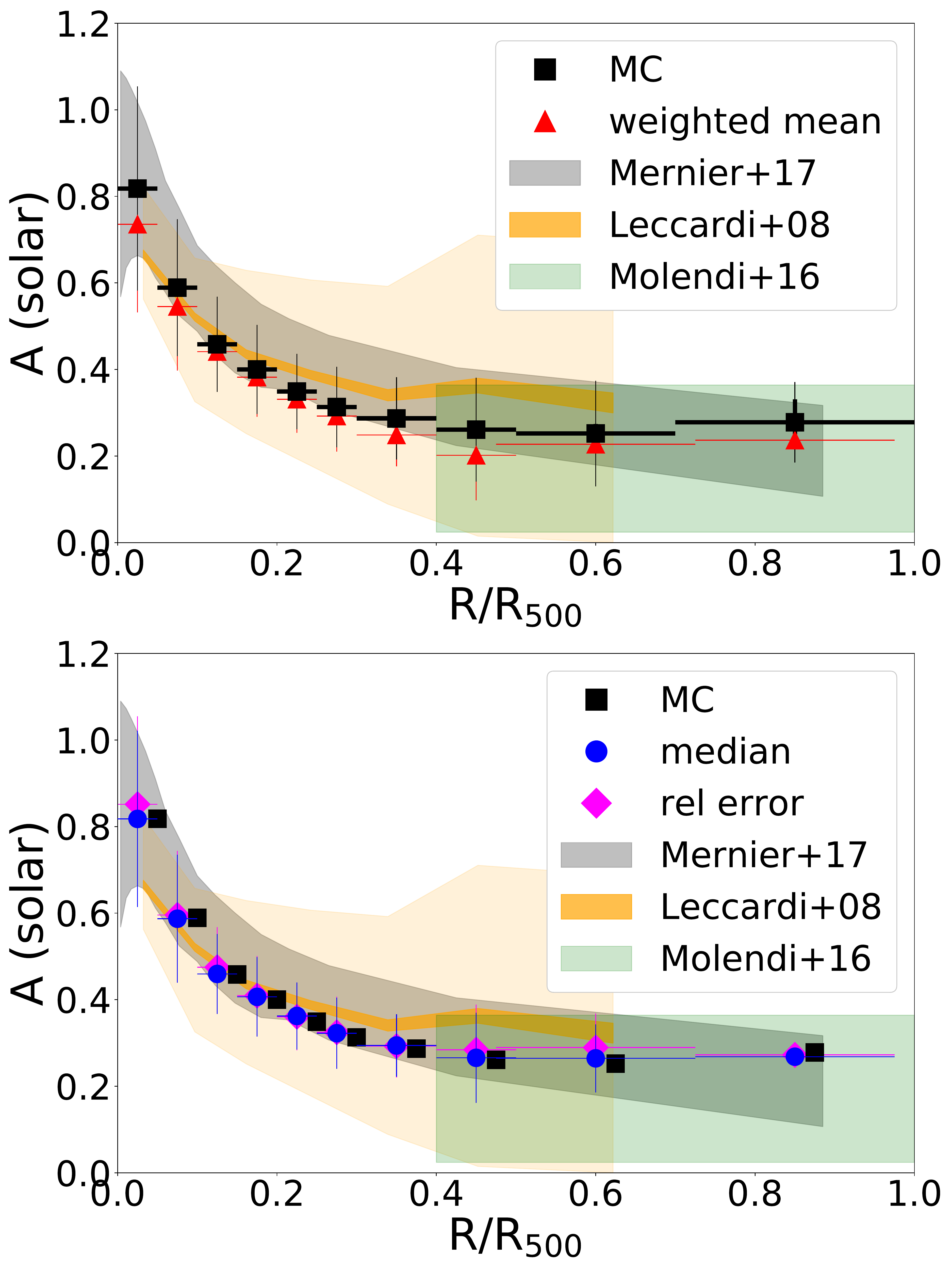}
\caption{{\it  top:}  stacked metallicity profiles for the most relaxed (bottom-left quadrant in Fig. 1) galaxy groups and clusters derived using the weighted mean (red triangles) and compared with the stacked profile derived with the MC simulations (black points).   ({\it bottom})  the same as in the left panel but comparing the profile obtained using the median (blue points) of the distribution with the weighted mean profile obtained using the relative errors as weights (magenta diamonds) and compared with the stacked profile derived with the MC simulations (black points, slightly shifted and without errorbars for visualization purposes).}
\end{figure}
The binned profiles presented in this paper have been obtained using Monte Carlo simulations instead of the weighted mean as done, e.g., by  \cite{2017A&A...603A..80M} and \cite{2008A&A...487..461L}. In Fig. D1 ({\it top panel}) we show the comparison between the stacked profile for the relaxed clusters obtained using MC simulations and the profile that one obtain following the prescription for the weighted mean (red triangles) given in \cite{2008A&A...487..461L}. The latter appears to be systematically lower at all radii. The reason is that some objects in our sample have a very good data quality leading to very small statistical uncertainties (in particular in the center where we set a minimum width of 30 arcsec per annulus) of their measured metallicities. As a consequence, these galaxy groups and clusters have an important contribution in the final average metallicity profile.  These long observations are often associated with particular classes of clusters, like strong cool cores or major mergers in contrast to the less exciting weak cool core clusters. Moreover, the quality of the data also depends on the fraction of flares and so the clean exposure time, introducing a bias, on top of the archival bias, that cannot be easily quantified.
When removing six clusters (i.e. A4038, Hydra-A, A0085, AS1101, 2A0335,  and A2199) that contribute to $\sim$65$\%$ of the total weight in the innermost annuli, this trend disappears and the two profiles overlap with only a residual $\sim$5$\%$ difference in the innermost bin. Those clusters are among the ones with the longest observations, and so delivering smallest statistical errors and so significantly contributing to the weighted mean values. Moreover, even with similar data quality clusters with lower metallicities have a smaller statistical errors and so biasing toward smaller values the average profile.  To avoid this problem one can use either the median values of each radial bin or use the relative errors instead of the statistical errors as weights. In Fig. D1 ({\it bottom panel}) we compare the profile obtained using the MC simulations with the profile obtained using the median values (blue points) with the weighted mean profile using the relative errors as weight (magenta diamonds). The agreement is good even without removing any cluster from the sample because the weight of every individual system is much smaller than in the case of the weight based on the inverse of the square of the individual errors.

\section{Cluster properties}
In Table C1 we provide the list of clusters (column 1) used in this paper. Coordinates and redshifts are shown in columns 2-4. In  columns 5-6 we provide the list of XMM-Newton observations which we investigated and the corresponding clean exposure times for MOS1, MOS2, and pn. The masses used to estimate $R_{500}$ are given in column 7 and are taken from \cite{2011A&A...534A.109P}. In columns 8-10 we provided the estimated core abundances, concentration, and centroid-shift parameters.

\begin{table*}
\renewcommand{\thetable}{C1}{\arabic{table}}
\centering
\caption{Properties of the clusters. }
\begin{tabular}{|c|ccc|ccc|ccc}
\hline
Name & RA & DEC & redshift  & ObsID & $t_{exp}$ & $M_{500}$ & A$_{core}$ & w & c  \\
& & & &  & [ks] & [$10^{14} M_{\odot}$] & [solar] &  & \\
\hline
NGC4936 & 13 04 16.7 & -30 30 55 & 0.012 & 0204540101 & 11.7, 12.9, 6.5 & 0.225 & 0.332$\pm$0.031 & 0.014$\pm$0.001 & 0.156$\pm$0.002 \\
A3565 &13 36 38.8 & -33 57 30 & 0.012 & 0672870101 & 30.6, 33.7, 19.4 & 0.128 & 0.450$\pm$0.045 & 0.017$\pm$0.001 & 0.194$\pm$0.002 \\
A1060 & 10 36 41.8 & -27 31 28  & 0.013 & 0206230101 & 31.4, 32.9, 23.9 & 0.994 & 0.471$\pm$0.007 & 0.009$\pm$0.001 & 0.181$\pm$0.000 \\
NGC1550 & 04 19 37.8 & +02 24 50  & 0.055 & 0152150101 & 19.0, 18.9, 14.8 & 0.703 & 0.549$\pm$0.004 & 0.004$\pm$0.000 & 0.352$\pm$0.001 \\
& & & & 0723800401 & 47.3, 48.5, 29.5 & & & & \\
& & & & 0723800501 & 79.2, 79.5, 59.6 & & & & \\
S0753 & 14 03 35.9 & -33 59 16 & 0.014 & 0741930101 & 108.5, 111.5, 89.7 & 0.267 & 0.209$\pm$0.008 & 0.010$\pm$0.001 & 0.229$\pm$0.001 \\
HCG62 & 12 53 05.5 & -09 12 01 & 0.015 & 0504780501 & 78.7, 79.5, 58.5 & 0.272 & 0.268$\pm$0.005 & 0.011$\pm$0.001 & 0.441$\pm$0.001 \\
NGC499 & 01 23 12.2 & +33 27 40 & 0.150 & 0402360101 & 20.0, 21.1, 10.6 & 0.362 & 0.429$\pm$0.010 & 0.009$\pm$0.001 & 0.439$\pm$0.002 \\
S0840 & 04 58 55.1 & -00 29 21  & 0.015 & 0673180401 & 19.0, 21.8, 12.6 & 0.114 & 0.306$\pm$0.052 & 0.019$\pm$0.001 & 0.376$\pm$0.007 \\
NGC3411 & 10 50 25.5 & -12 50 47 & 0.015 & 0146510301 & 20.1, 20.8, 17.3 & 0.411 & 0.776$\pm$0.021 & 0.005$\pm$0.000 & 0.623$\pm$0.003 \\
A0262 & 01 52 46.8 & +36 09 05 & 0.017 & 0109980101 & 21.3, 21.4, 15.5 & 1.189 & 0.485$\pm$0.007 & 0.005$\pm$0.000 & 0.245$\pm$0.001 \\
& & & & 0504780201 & 25.9, 28.4, 14.5 & & & & \\
NGC507 & 01 23 41.0 & +33 15 40 & 0.017 & 0723800301 & 80.5, 83.2, 60.0 & 0.611 & 0.432$\pm$0.004 & 0.033$\pm$0.002 & 0.325$\pm$0.001 \\
NGC777 & 02 00 16.5 & +31 26 11 & 0.170 & 0203610301 & 5.2, 5.2, 1.6 & 0.285 & 0.819$\pm$0.142 & 0.007$\pm$0.000 & 0.281$\pm$0.003 \\
& & & & 0304160301 & 4.9, 6.3, 1.6 & & & & \\
AWM7 & 02 54 29.5 & +41 34 18  &  0.017 & 0135950301 & 27.4, 27.8, 22.0 & 1.835 & 0.642$\pm$0.003 & 0.017$\pm$0.001 & 0.199$\pm$0.000 \\
& & & & 0605540101 & 109.2, 112.2, 83.7 & & & & \\
MKW1s & 09 20 00.5 & +01 02 24 & 0.018 & 0673180201 & 14.8, 16.7, 5.3 & 0.134 & 0.595$\pm$0.338 & 0.017$\pm$0.001 & 0.254$\pm$0.005 \\
NGC0410 & 01 10 58.1 & +33 08 58 & 0.015 & 0203610201 & 13.2, 12.9, 10.6 & 0.215 & 0.297$\pm$0.022 & 0.012$\pm$0.001 & 0.420$\pm$0.005 \\
A0189 & 01 25 24.7 & +01 44 28 & 0.033 & 0109860101 & 32.9, 34.3, 27.3 & 0.370 & 0.473$\pm$0.012 & 0.004$\pm$0.000 & 0.380$\pm$0.002 \\
MKW4 & 12 04 25.2 & +01 54 02 & 0.020 & 0723800601 & 15.3, 17.4, 8.4 & 0.677 & 0.768$\pm$0.010 & 0.005$\pm$0.000 & 0.348$\pm$0.001 \\
& & & & 0723800701 & 54.5, 55.7, 34.4 & & & & \\
HCG97 & 23 47 24.4 & -02 18 52 & 0.022 & 0152860101 & 24.5, 24.0, 20.7 & 0.245 & 0.342$\pm$0.021 & 0.026$\pm$0.001 & 0.351$\pm$0.004 \\
NGC5171 & 13 29 27.8 & +11 43 23 & 0.023 & 0041180801 & 16.1, 15.7, 11.4 & 0.170 & 0.225$\pm$0.034 & 0.026$\pm$0.001 & 0.087$\pm$0.002 \\
S0301 & 02 49 36.9 & -31 11 19 & 0.022 & 0146510401 & 29.9, 29.4, 23.0 & 0.447 & 0.494$\pm$0.010 & 0.010$\pm$0.001 & 0.414$\pm$0.002 \\
A3581 & 14 07 28.1 & -27 00 55 & 0.023 & 0205990101 & 33.1, 33.7, 28.2 & 1.081 & 0.457$\pm$0.004 & 0.008$\pm$0.000 & 0.470$\pm$0.001 \\
& & & & 0504780301 & 41.5, 47.2, 26.3 & & & & \\
NGC1132 & 02 52 49.4 & -01 16 27 & 0.024 & 0151490101 & 19.7, 20.4, 15.0 & 0.466 & 0.047$\pm$0.041 & 0.003$\pm$0.000 & 0.280$\pm$0.003 \\
A2877 & 01 10 00.4 & -45 55 22 & 0.025 & 0204540201 & 19.2, 19.3, 15.2 & 0.710 & 0.420$\pm$0.027 & 0.039$\pm$0.002 & 0.155$\pm$0.001 \\
A0400 & 02 57 38.9 & +06 00 22 & 0.024 & 0404010101 & 25.6, 26.0, 16.7  & 0.801 & 0.597$\pm$0.021 & 0.037$\pm$0.002 & 0.152$\pm$0.001 \\
Zw1745 & 17 36 22.1 & +68 03 26 & 0.025 & 0203610401 & 20.8, 21.2, 15.7 & 0.239 & 0.261$\pm$0.027 & 0.063$\pm$0.004 & 0.178$\pm$0.004 \\
IC1867 & 02 55 51.3 & +09 18 48 & 0.026 & 0203610501 & 12.1, 11.8, 4.5 & 0.325 & 0.748$\pm$0.110 & 0.005$\pm$0.000 & 0.358$\pm$0.005 \\
NGC4325 & 12 23 06.5 & +10 37 26 & 0.026 & 0108860101 & 17.9, 17.6, 13.2 & 0.559 & 0.570$\pm$0.013 & 0.002$\pm$0.000 & 0.358$\pm$0.005 \\
MKW8 & 14 40 38.2 & +03 28 35 & 0.026 & 0300210701 & 19.7, 20.3, 14.5 & 0.735 & 0.395$\pm$0.023 & 0.021$\pm$0.001 & 0.153$\pm$0.001 \\
R2315.7--0222 & 23 15 45.2 & -02 22 37 & 0.025 & 0501110101 & 30.8, 30.3, 24.2 & 0.584 & 0.535$\pm$0.018 & 0.019$\pm$0.001 & 0.330$\pm$0.002 \\
Ophiucus & 17 12 24.7 & -23 21 01  &  0.055 & 0505150101 & 27.4, 29.0, 14.9 & 5.312 & 0.483$\pm$0.008 & 0.033$\pm$0.002 & 0.133$\pm$0.000 \\
NGC4104 & 12 06 37.4 & +28 11 01 & 0.028 & 0301900401 & 11.1, 11.1, 8.4 & 0.418 & 0.269$\pm$0.021 & 0.006$\pm$0.000 & 0.250$\pm$0.004 \\
A2199 & 16 28 38.0 & +39 32 55 & 0.030 & 0723801101 & 46.2, 47.7, 36.3 & 2.963 & 0.506$\pm$0.003 & 0.008$\pm$0.001 & 0.310$\pm$0.000 \\
& & & & 0723801201 & 47.1, 47.6, 39.5 & & & & \\
A4038 & 23 47 43.2 & -28 08 29 & 0.028 & 0204460101 & 26.6, 25.7, 22.8 & 2.038 & 0.436$\pm$0.005 & 0.006$\pm$0.000 & 0.405$\pm$0.001 \\
& & & & 0723800801 & 44.0, 44.1, 36.3 & & & & \\
R0953.2--1558 & 09 53 12.1 & -15 58 52 & 0.030 & 0140210201 & 34.2, 34.8, 28.1 & 0.418 & 0.400$\pm$0.015 & 0.009$\pm$0.001 & 0.385$\pm$0.003 \\
IC1262 & 17 33 02.6 & +43 45 46  & 0.031 & 0741580201 & 5.9, 6.0, 3.2 & 0.859 & 0.312$\pm$0.018 & 0.011$\pm$0.001 & 0.358$\pm$0.004 \\
A2634 & 23 38 25.7 & +27 00 45 & 0.031 & 0002960101 & 7.1, 7.4, 3.8 & 1.215 & 0.339$\pm$0.038 & 0.002$\pm$0.000 & 0.061$\pm$0.001 \\
& & & & 0505210801 & 6.8, 8.9, 2.7 & & & & \\
IIIZw54 & 03 41 16.9 & +15 24 27 & 0.031 & 0505230401 & 35.2, 38.4, 23.4 & 1.130 & 0.263$\pm$0.011 & 0.004$\pm$0.000 & 0.277$\pm$0.001 \\
A0496 & 04 33 38.4 & -13 15 33 & 0.033 & 0506260301 & 51.1, 53.0, 36.3 & 2.912 & 0.656$\pm$0.004 & 0.009$\pm$0.001 & 0.430$\pm$0.001 \\
& & & & 0506260401 & 50.5, 51.5, 39.0 & & & & \\
AWM4 & 16 04 57.0 & +23 55 14   & 0.032 & 0093060401 & 15.3, 15.4, 10.9 & 0.929 & 0.511$\pm$0.022 & 0.002$\pm$0.000 & 0.315$\pm$0.002 \\ 
CID28 & 04 54 50.3 & -18 06 33 & 0.034 & 0140210101 & 29.0, 28.5, 23.9 & 0.619 & 0.573$\pm$0.027 & 0.009$\pm$0.001 & 0.240$\pm$0.002 \\
AWM5 & 16 58 00.8 & +27 51 16 & 0.034 & 0654800201 & 44.5, 44.1, 37.4 & 0.708 & 0.502$\pm$0.027 & 0.010$\pm$0.001 & 0.282$\pm$0.002 \\
& & & & 0670350701 & 9.1, 8.9, 6.3 & & & & \\
UGC03957 & 07 40 59.4 & +55 25 55 & 0.034 & 0653580101 & 23.5, 24.7, 8.2 & 1.290 & 0.673$\pm$0.019 & 0.008$\pm$0.000 & 0.456$\pm$0.002 \\
A1314 & 11 34 50.5 & +49 03 28 & 0.034 & 0149900201 & 16.1, 16.0, 13.1 & 0.461 & 0.258$\pm$0.029 & 0.025$\pm$0.001 & 0.085$\pm$0.002 \\
IC1880 & 03 06 28.7 & -09 43 50 & 0.034 & 0601930401 & 56.1, 55.5, 34.0 & 0.289 & 0.354$\pm$0.016 & 0.004$\pm$0.000 & 0.452$\pm$0.004 \\
2A0335 & 03 38 40.8 & +09 58 28 & 0.035 & 0147800201 & 95.1, 96.6, 84.9 & 3.450 & 0.574$\pm$0.003 & 0.007$\pm$0.000 & 0.614$\pm$0.001 \\
R0340.6--0239 & 03 40 41.8 & -02 39 57 & 0.035 & 0741580901 & 6.3, 7.0, 3.8 & 0.795 & 0.450$\pm$0.034 & 0.009$\pm$0.001 & 0.390$\pm$0.007 \\
A2052 & 15 16 44.0 & +07 01 07 & 0.035 & 0109920101 & 26.7, 27.0, 20.6 & 2.494 & 0.569$\pm$0.007 & 0.002$\pm$0.000 & 0.425$\pm$0.001 \\
& & & & 0401521201 & 14.9, 15.2, 11.1 & & & & \\
A2147 & 16 02 18.7 & +16 01 12 & 0.035 & 0505210601 & 9.4, 9.1, 4.6 & 2.405 & 0.363$\pm$0.036 & 0.014$\pm$0.001 & 0.120$\pm$0.001 \\
\hline
\end{tabular}
\end{table*}

\begin{table*}
 \contcaption{}
\begin{tabular}{|c|ccc|ccc|ccc}
\hline
Name & RA & DEC & redshift  & ObsID & $t_{exp}$ & $M_{500}$ & A$_{core}$ & w & c  \\
& & & &  & [ks] & [$10^{14} M_{\odot}$] & [solar] &  & \\
\hline
A2063 & 15 23 05.4 & +08 36 09 & 0.035 & 0200120401 & 1.3, 6.6, - & 2.160 & 0.491$\pm$0.014 & 0.004$\pm$0.000 & 0.258$\pm$0.001 \\
& & & & 0550360101 & 18.9, 21.4, 11.4 & & & & \\
S0540 & 05 40 06.3 & -40 50 32 & 0.036 & 0149420101 & 10.7, 11.6, 5.2 & 1.221 & 0.528$\pm$0.031 & 0.005$\pm$0.000 & 0.313$\pm$0.003 \\
NGC1650 & 04 45 10.0 & -15 51 01 & 0.036 & 0741580701 & 4.9, 5.0, 3.9 & 0.864 & 0.526$\pm$0.078 & 0.019$\pm$0.001 & 0.162$\pm$0.004 \\
NGC5098 & 13 20 15.4 & +33 08 30 & 0.036 & 0105860101 & 29.8, 30.0, 23.3 & 0.517 & 0.341$\pm$0.013 & 0.009$\pm$0.001 & 0.377$\pm$0.003 \\
A2151 & 16 04 35.7 & +17 43 28 & 0.037 & 0147210301 & 7.1, 7.4, 4.5 & 1.320 & 0.352$\pm$0.022 & 0.018$\pm$0.001 & 0.314$\pm$0.003 \\
R1423.1+2615 & 14 23 10.1 & +26 15 20 & 0.037 & 0670350101 & 10.0, 10.9, 4.9 & 0.376  & 0.528$\pm$0.040 & 0.020$\pm$0.001 & 0.408$\pm$0.007 \\
A3570 & 13 46 52.5 & -37 52 28 & 0.038 & 0765030701 & 16.9, 18.0, 11.7 & 0.699 & 0.546$\pm$0.073 & 0.130$\pm$0.007 & 0.050$\pm$0.002 \\
A0576 & 07 21 22.2 & +55 47 11 & 0.039 & 0205070301 & 9.0, 9.7, 6.5 & 1.681 & 0.475$\pm$0.011 & 0.014$\pm$0.001 & 0.206$\pm$0.001 \\
& & & & 0205070401 & 13.8, 14.5, 10.6 & & & & \\
& & & & 0504320201 & 20.8, 21.7, 14.0 & & & & \\
& & & & 0504320101 & 25.9, 28.4, 16.7 & & & & \\
A1139 & 10 58 10.4 & +01 35 11 & 0.037 & 0601930101 & 24.8, 27.5, 17.5 & 0.439 & 0.407$\pm$0.045 & 0.020$\pm$0.001 & 0.087$\pm$0.002 \\
CID36 & 05 42 09.3 & -26 07 25 & 0.039 & 0741581101 & 10.7, 11.6, 7.9 & 0.853 & 0.362$\pm$0.034 & 0.035$\pm$0.002 & 0.154$\pm$0.003 \\
A3571 & 13 47 28.4 & -32 50 59 & 0.039 & 0086950201 & 19.7, 19.5, 13.1  & 4.507 & 0.478$\pm$0.014 & 0.011$\pm$0.001 & 0.258$\pm$0.001 \\
RBS0540 & 04 25 51.4 & -08 33 33 & 0.040 & 0300210401 & 36.0, 35.9, 27.3 & 2.086 & 0.523$\pm$0.010 & 0.003$\pm$0.000 & 0.525$\pm$0.001 \\
R0748.1+1832 & 07 48 09.5 & +18 32 47 & 0.040 & 0651780201 & 12.2, 11.9, 7.1  & 0.520 & 0.404$\pm$0.053 & 0.003$\pm$0.000 & 0.159$\pm$0.003 \\
R0137.2--0912 & 01 37 15.4 & -09 12 10 & 0.041 & 0765001101 & 5.7, 5.6, 3.4  & 0.947 & 0.360$\pm$0.024 & 0.011$\pm$0.001 & 0.367$\pm$0.005 \\
A2589 & 23 23 53.5 & +16 48 32 & 0.042 & 0204180101 & 22.1, 23.1, 17.9 & 1.990 & 0.591$\pm$0.017 & 0.006$\pm$0.000 & 0.284$\pm$0.001 \\
R1742.8+3900 & 17 42 48.3 & +39 00 35 & 0.042 & 0765040801 & 12.7, 12.9, 11.2 & 0.774 & 0.208$\pm$0.009 & 0.012$\pm$0.001 & 0.379$\pm$0.004 \\
A0548 & 05 45 27.2 & -25 56 20 & 0.042 & 0302030101 & 42.7, 42.9, 29.2 & 0.502 & 0.636$\pm$0.060 & 0.012$\pm$0.001 & 0.139$\pm$0.002 \\
R1740.5+3538 & 17 40 32.7 & +35 38 51 & 0.043 & 0761112101 & 16.5, 16.1, 12.4  & 0.914 & 0.599$\pm$0.027 & 0.010$\pm$0.001 & 0.344$\pm$0.003 \\
A0119 & 00 56 18.3 & -01 13 00 & 0.044 & 0012440101 & 22.1, 22.7, 16.2 & 2.475 & 0.338$\pm$0.024 & 0.003$\pm$0.000 & 0.090$\pm$0.001 \\
& & & & 0505211001 & 9.1, 9.7, 6.3 & & & & \\
A0160 & 01 13 05.8 & +15 31 01 & 0.044 & 0651780101 & 11.1, 12.6, 4.9 & 0.745 & 0.462$\pm$0.080 & 0.027$\pm$0.002 & 0.136$\pm$0.003 \\
S0555 & 05 57 13.2 & -37 27 58  &  0.044 & 0765010701 & 14.7, 14.7, 12.3 & 0.972 & 0.495$\pm$0.024 & 0.003$\pm$0.000 & 0.391$\pm$0.004 \\
MKW3s & 15 21 50.0 & +07 42 32 & 0.044 & 0723801501 & 100.1, 101.4, 77.9 & 2.522 & 0.529$\pm$0.006 & 0.006$\pm$0.000 & 0.413$\pm$0.001 \\
A1983 & 14 52 58.8 & +16 41 59 & 0.044 & 0091140201 & 23.3, 21.9, 12.5 & 0.878 & 0.525$\pm$0.037 & 0.005$\pm$0.000 & 0.209$\pm$0.002 \\
RBS0485 & 03 52 20.7 & -54 53 09  & 0.045 & 0651580601 & 22.3, 23.3, 11.4 & 0.681 & 0.539$\pm$0.044 & 0.018$\pm$0.001 & 0.157$\pm$0.003 \\
A3558C & 13 31 32.4 & -31 48 55 & 0.045 & 0105261401 & 10.2, 10.5, 6.6 & 0.833 & 0.288$\pm$0.057 & 0.013$\pm$0.001 & 0.114$\pm$0.002 \\
R1454.4+1622 & 14 54 28.0 & +16 22 13 & 0.046 & 0670350201 & 15.4, 16.4, 14.9 & 0.461 & 0.243$\pm$0.031 & 0.003$\pm$0.000 & 0.178$\pm$0.004 \\
A1736 & 13 26 54.0 & -27 11 00 & 0.046 & 0505210201 & 10.1, 9.4, 3.6 & 2.706 & 0.421$\pm$0.044 & 0.042$\pm$0.002 & 0.095$\pm$0.001 \\
A3376 & 06 01 45.7 & -39 59 34 & 0.047 & 0151900101 & 21.6, 22.0, 15.3 & 1.976 & 0.362$\pm$0.021 & 0.082$\pm$0.005 & 0.109$\pm$0.001 \\
& & & & 0504140101 & 36.6, 39.0, 28.1 & & & & \\
A1644 & 12 57 09.7 & -17 24 01 & 0.047 & 0010420201 & 12.4, 12.8, 10.0 & 2.925 & 0.464$\pm$0.027 & 0.012$\pm$0.001 & 0.189$\pm$0.001 \\
A4059 & 23 57 02.3 & -34 45 38 & 0.048 & 0723800901 & 66.9, 68.9, 40.6 & 2.666 & 0.627$\pm$0.006 & 0.004$\pm$0.000 & 0.363$\pm$0.001 \\
& & & & 0723801001 & 70.6, 70.5, 52.8 & & & & \\
A3558 & 13 27 57.5 & -31 30 09  &  0.048 & 0107260101 & 38.9, 39.4, 32.4 & 3.974 & 0.420$\pm$0.010 & 0.027$\pm$0.002 & 0.200$\pm$0.001 \\
A3560 & 13 32 22.6 & -33 08 22 & 0.049 & 0205450201 & 26.9, 27.4, 17.5 & 1.682 & 0.396$\pm$0.030 & 0.021$\pm$0.001 & 0.126$\pm$0.001 \\
A3558B & 13 29 42.9 & -31 36 09 & 0.049 & 0651590201 & 8.5, 9.7, - & 1.721 & 0.370$\pm$0.053 & 0.005$\pm$0.000 & 0.173$\pm$0.003 \\
A2717 & 00 03 12.1 & -35 55 38 & 0.049 & 0145020201 & 46.3, 46.1, 40.2 & 1.202 & 0.588$\pm$0.019 & 0.006$\pm$0.000 & 0.297$\pm$0.001 \\
A3562 & 13 33 36.3 & -31 39 40 & 0.049 & 0105261501 & 15.6, 17.0, 8.3 & 2.370 & 0.480$\pm$0.015 & 0.008$\pm$0.000 & 0.222$\pm$0.001 \\
& & & & 0105261601 & 16.2, 15.8, 13.4 & & & & \\
& & & & 0105261301 & 34.9, 35.1, 31.1 & & & & \\
R1022.0+3830 & 10 22 04.7 & +38 30 43  &  0.049 & 0503601301 & 10.5, 10.1, 7.9 & 0.701 & 0.384$\pm$0.047 & 0.027$\pm$0.002 & 0.325$\pm$0.006 \\
R2104.9--5149 & 21 04 54.7 & -51 49 35 & 0.049 & 0765000101 & 10.7, 10.6, 6.2 & 1.323 & 0.608$\pm$0.034 & 0.004$\pm$0.000 & 0.414$\pm$0.004 \\
R1002.6+3241 & 10 02 38.6 & +32 41 58 & 0.050 & 0503600301 & 8.2, 7.7, 5.7 & 0.944 & 0.348$\pm$0.050 & 0.006$\pm$0.000 & 0.189$\pm$0.005 \\
R0413.9--3806 & 04 13 57.1 & -38 06 00 & 0.050 & 0720251501 & 8.2, 9.1, 4.0 & 1.625 & 0.553$\pm$0.022 & 0.005$\pm$0.000 & 0.323$\pm$0.003 \\
& & & & 0720253501 & 11.9, 12.4, 3.5 & & & & \\
A3391 & 06 26 22.8 & -53 41 44 & 0.051 & 0505210401 & 22.2, 21.9, 14.4 & 2.161 & 0.400$\pm$0.022 & 0.011$\pm$0.001 & 0.149$\pm$0.001 \\
& & & & 0720252301 & 11.0, 11.0, 8.7 & & & & \\
R1337.4--4120 & 13 37 28.2 & -41 20 01 & 0.052 & 0765041301 & 15.7, 16.9, 9.6 &  0.987 & 0.417$\pm$0.045 & 0.008$\pm$0.000 & 0.160$\pm$0.002 \\
A0151N & 01 08 50.1 & -15 24 36 & 0.053 & 0765001201 & 6.6, 8.7, - & 1.278 & 0.513$\pm$0.063 & 0.004$\pm$0.000 & 0.221$\pm$0.004 \\
A3301 & 05 00 46.5 & -38 40 41 & 0.054 & 0083151201 & 9.7, 9.8, 5.9 & 1.2705 & 0.319$\pm$0.061 & 0.013$\pm$0.001 & 0.161$\pm$0.003 \\
HYDRA & 09 18 06.5 & -12 05 36 & 0.054 & 0504260101 & 63.9, 67.6, 44.2 & 3.624 & 0.387$\pm$0.005 & 0.003$\pm$0.000 & 0.494$\pm$0.001 \\
R0631.3--5610 & 06 31 20.7 & -56 10 20 & 0.054 & 0720253401 & 23.2, 24.3, 19.4 & 1.286 & 0.326$\pm$0.037 & 0.030$\pm$0.002 & 0.127$\pm$0.002 \\
R0844.9+4258 & 08 44 56.7 & +42 58 54 & 0.055 & 0503600101 & 15.5, 16.4, 10.9 & 0.470 & 0.447$\pm$0.041 & 0.013$\pm$0.001 & 0.348$\pm$0.006 \\
A3530 & 12 55 34.5 & -30 19 50 & 0.054 & 0201780101 & 11.0, 10.7, 8.4 & 1.558 & 0.372$\pm$0.046 & 0.011$\pm$0.001 & 0.131$\pm$0.002 \\
  \hline
 \end{tabular}
\end{table*}

\begin{table*}
 \contcaption{}
\begin{tabular}{|c|ccc|ccc|ccc}
\hline
Name & RA & DEC & redshift  & ObsID & $t_{exp}$ & $M_{500}$ & A$_{core}$ & w & c  \\
& & & &  & [ks] & [$10^{14} M_{\odot}$] & [solar] &  & \\
\hline
A0754 & 09 09 08.4 & -09 39 58 & 0.054 & 0556200101 & 36.5, 37.4, 21.9 & 4.482 & 0.387$\pm$0.013 & 0.079$\pm$0.004 & 0.143$\pm$0.000 \\
& & & & 0556200501 & 68.6, 70.7, 49.1 & & & & \\
A3532 & 12 57 16.9 & -30 22 37 & 0.055 & 0030140301 & 7.9, 8.3, 4.6 & 2.335 & 0.412$\pm$0.051 & 0.038$\pm$0.002 & 0.117$\pm$0.002 \\
A0085 & 00 41 50.1 & -09 18 07 & 0.055 & 0723802101 & 82.2, 85.5, 55.8 & 5.316 & 0.562$\pm$0.004 & 0.009$\pm$0.001 & 0.364$\pm$0.000 \\
& & & & 0723802201 & 88.7, 91.3, 73.5 & & & & \\
A2319 & 19 21 08.8 & +43 57 30 & 0.056 & 0302150101 & 14.6, 14.6, 9.2 & 5.835 & 0.382$\pm$0.008 & 0.025$\pm$0.001 & 0.204$\pm$0.000 \\
& & & & 0302150201 & 14.3, 14.3, 9.8 & & & & \\
& & & & 0600040101 & 46.4, 45.4, 38.9 & & & & \\
R1122.2+6712 & 11 22 14.5 & +67 12 46 & 0.056 & 0503600401 & 14.9, 14.8, 10.9 & 0.439 & 0.568$\pm$0.063 & 0.008$\pm$0.001 & 0.359$\pm$0.007 \\
R1256.9--3119 & 12 56 59.8 & -31 19 19 & 0.056 & 0765040401 & 5.7, 6.9, 3.1 & 1.108 & 0.551$\pm$0.133 & 0.005$\pm$0.000 & 0.182$\pm$0.006 \\
S0868 & 20 23 01.6 & -20 56 55 & 0.056 & 0201902301 & 14.9, 15.4, 7.9 & 1.182 & 0.293$\pm$0.040 & 0.018$\pm$0.001 & 0.197$\pm$0.003 \\
Sersic & 23 13 58.6 & -42 44 02 & 0.058 & 0147800101 & 78.8, 78.6, 64.0 & 2.826 & 0.406$\pm$0.005 & 0.005$\pm$0.000 & 0.577$\pm$0.001 \\
A2626 & 23 36 30.3 & +21 08 33 & 0.055 & 0083150201 & 8.7, 8.1, 4.4 & 1.806 & 0.602$\pm$0.013 & 0.006$\pm$0.000 & 0.361$\pm$0.002 \\
& & & & 0148310101 & 35.5, 36.8, 29.8 & & & & \\
A0133 & 01 02 42.1 & -21 52 25 & 0.057 & 0144310101 & 18.6, 18.8, 13.8 & 2.477 & 0.694$\pm$0.008 & 0.002$\pm$0.000 & 0.402$\pm$0.001 \\
& & & & 0723801301 & 62.3, 63.5, 49.6 & & & & \\
& & & & 0723802001 & 29.0, 33.2, 14.2 & & & & \\
R1926.9--5342 & 19 26 58.3 & -53 42 11 & 0.057 & 0765000601 & 13.4, 14.9, 7.5 & 1.489 & 0.305$\pm$0.016 & 0.007$\pm$0.000 & 0.468$\pm$0.004 \\
A2401 & 21 58 20.1 & -20 06 16 & 0.057 & 0555220101 & 50.5, 51.5, 40.3 & 0.898 & 0.338$\pm$0.032 & 0.003$\pm$0.000 & 0.175$\pm$0.002 \\
R0746.3+3100 & 07 46 37.3 & +31 00 49 & 0.058 & 0503600201 & 10.4, 10.5, 7.6 & 0.675 & 0.522$\pm$0.073 & 0.009$\pm$0.001 & 0.170$\pm$0.004 \\
A0152 & 01 10 05.5 & +13 58 49 & 0.058 & 0503600701 & 13.2, 13.7, 6.3 & 0.408 & 0.435$\pm$0.083 & 0.009$\pm$0.001 & 0.261$\pm$0.007 \\
A1991 & 14 54 31.4 & +18 38 31 & 0.059 & 0145020101 & 23.5, 24.0, 14.5 & 1.676 & 0.594$\pm$0.016 & 0.002$\pm$0.000 & 0.462$\pm$0.002 \\
R2124.3--7446 & 21 24 22.8 & -74 46 25 & 0.059 & 0765040101 & 11.5, 12.3, 7.7 & 1.174 & 0.646$\pm$0.050 & 0.006$\pm$0.000 & 0.347$\pm$0.004 \\
A3266 & 04 31 24.1 & -61 26 38  & 0.059 & 0105261101 & 11.3, 10.8, 6.0 & 4.558 & 0.303$\pm$0.016 & 0.059$\pm$0.003 & 0.133$\pm$0.000 \\
& & & & 0105260701 & 18.0, 18.3, 12.8 & & & & \\
& & & & 0105260901 & 20.9, 20.9, 14.2 & & & & \\
& & & & 0105260801 & 18.6, 17.8, 12.5 & & & & \\
A3158 & 03 42 53.9 & -53 38 07 & 0.059 & 0300210201 & 17.7, 17.4, 9.1 & 3.651 & 0.431$\pm$0.016 & 0.005$\pm$0.000 & 0.202$\pm$0.001 \\
& & & & 0300211301 & 7.4, 7.6, 4.0 & & & & \\
S0974 & 21 47 55.5 & -46 00 19 & 0.059 & 0765010801 & 11.1, 12.3, 5.1 & 1.376 & 0.268$\pm$0.028 & 0.020$\pm$0.001 & 0.260$\pm$0.004 \\
A3651 & 19 52 16.5 & -55 03 42 & 0.060 & 0720252001 & 3.9, 3.8, 2.4 & 1.475 & 0.383$\pm$0.045 & 0.015$\pm$0.001 & 0.111$\pm$0.002 \\
S0384 & 03 45 45.7 & -41 12 27 & 0.060 & 0201900801 & 11.0, 12.5, 2.1 & 1.270 & 0.632$\pm$0.025 & 0.006$\pm$0.000 & 0.418$\pm$0.004 \\
& & & & 0404910301 & 12.9, 13.6, 6.1 & & & & \\
R0225.1--2928 & 02 25 10.5 & -29 28 26 & 0.060 & 0302610601 & 14.7, 15.2, 7.2 & 1.121 & 0.678$\pm$0.055 & 0.017$\pm$0.001 & 0.259$\pm$0.004 \\
S0405 & 03 51 08.9 & -82 13 00 & 0.061 & 0675471101 & 6.4, 7.4, 1.9 & 2.192 & 0.317$\pm$0.034 & 0.016$\pm$0.001 & 0.163$\pm$0.002 \\
& & & & 0720250601 & 3.7, 10.3, 6.1  & & & & \\
A2622 & 23 35 05.0 & +27 22 12 & 0.061 & 0765020701 & 19.2, 18.9, 14.1 & 1.336 & 0.577$\pm$0.038 & 0.004$\pm$0.000 & 0.373$\pm$0.004 \\
A2734 & 00 11 20.7 & -28 51 18 & 0.062 & 0675470801 & 9.4, 10.1, 6.3 & 2.061 & 0.419$\pm$0.042 & 0.016$\pm$0.001 & 0.204$\pm$0.002 \\
A0602 & 07 53 24.2 & +29 21 58 & 0.062 & 0761112401 & 9.2, 9.3, 7.0 & 1.424 & 0.397$\pm$0.048 & 0.048$\pm$0.003 & 0.122$\pm$0.002 \\
A1795 & 13 48 53.0 & +26 35 44 & 0.062 & 0097820101 & 34.3, 32.5, 23.3 & 5.528 & 0.471$\pm$0.008 & 0.006$\pm$0.000 & 0.471$\pm$0.001 \\
S0239 & 02 16 42.3 & -47 49 24 & 0.064 & 0501110201 & 38.3, 38.2, 27.7 & 1.008 & 0.479$\pm$0.034 & 0.003$\pm$0.000 & 0.214$\pm$0.003 \\
A3122 & 03 22 18.6 & -41 21 34 & 0.064 & 0720253201 & 21.4, 22.1, 17.1 & 1.424 & 0.325$\pm$0.044 & 0.045$\pm$0.003 & 0.115$\pm$0.002 \\
S1136 &  23 36 17.0 & -31 36 37 & 0.064 & 0765041001 & 9.8, 10.2, 6.8 & 1.289 & 0.560$\pm$0.073 & 0.004$\pm$0.000 & 0.153$\pm$0.003 \\
R2306.8--1324 & 23 06 51.7 & -13 24 59 & 0.066 & 0765030201 & 7.9, 9.0, 5.5 & 1.085 & 0.330$\pm$0.065 & 0.028$\pm$0.002 & 0.345$\pm$0.008 \\
S0112 & 00 57 48.1 & -66 48 44 & 0.067 & 0653880201 & 31.2, 33.4, 20.4 & 1.620 & 0.360$\pm$0.039 & 0.062$\pm$0.003 & 0.152$\pm$0.002 \\
A0500 & 04 38 54.7 & -22 06 49 & 0.067 & 0720253301 & 29.7, 28.7, 24.9 & 1.509 & 0.510$\pm$0.037 & 0.008$\pm$0.001 & 0.142$\pm$0.002 \\
A3497 & 12 00 05.0 & -31 24 21 & 0.069 & 0761112801 & 7.2, 7.6, 4.8 & 1.565 & 0.352$\pm$0.046 & 0.042$\pm$0.002 & 0.212$\pm$0.004 \\
S0987 & 22 01 50.9 & -22 26 40 & 0.069 & 0765030901 & 13.7, 13.7, 10.8 & 1.457 & 0.212$\pm$0.046 & 0.005$\pm$0.000 & 0.208$\pm$0.004 \\
A3490 & 11 45 19.1 & -34 25 43 & 0.070 & 0720252801 & 12.3, 11.0, 10.0 & 1.640 & 0.420$\pm$0.046 & 0.007$\pm$0.000 & 0.154$\pm$0.002 \\
A1837 & 14 01 36.7 & -11 07 28 & 0.070 & 0109910101 & 43.7, 43.5, 35.8 & 1.482  & 0.517$\pm$0.024 & 0.013$\pm$0.001 & 0.194$\pm$0.001 \\
Z1420+4933 & 14 21 35.5 & +49 33 07 & 0.072 & 0765020201 & 8.8, 8.3, 6.8 & 1.687 & 0.422$\pm$0.031 & 0.011$\pm$0.001 & 0.358$\pm$0.004 \\
A3104 & 03 14 19.8 & -45 25 27 & 0.072 & 0765000401 & 12.7, 12.6, 10.1 & 1.979 & 0.632$\pm$0.031 & 0.009$\pm$0.001 & 0.370$\pm$0.003 \\
A0399 & 02 57 49.8 & +13 02 57 & 0.072 & 0112260101 & 10.6, 10.8, 5.4 & 4.245 & 0.253$\pm$0.030 & 0.046$\pm$0.003 & 0.164$\pm$0.002 \\
A2065 & 15 22 26.5 & +27 42 34 & 0.072 & 0112240201 & 9.9, 9.2, 6.2 & 3.508 & 0.388$\pm$0.018 & 0.033$\pm$0.002 & 0.234$\pm$0.001 \\
& & & & 0202080201 & 18.4, 18.2, 12.9 & & & & \\
A1775 & 13 41 53.8 & +26 22 19 & 0.072 & 0108460101 & 21.0, 20.9, 16.2 & 2.478 & 0.608$\pm$0.028 & 0.007$\pm$0.000 & 0.412$\pm$0.002 \\
S0810 & 19 12 40.3 & -75 17 30 & 0.073 & 0720251001 & 15.3, 15.0, 12.1 & 2.154 & 0.467$\pm$0.030 & 0.020$\pm$0.001 & 0.309$\pm$0.003 \\
R1539.5--8335 & 15 39 33.9 & -83 35 32 & 0.073 & 0720252501 & 10.1, 10.4, 8.1 & 3.177 & 0.625$\pm$0.025 & 0.002$\pm$0.000 & 0.606$\pm$0.004 \\
A1589 & 12 41 19.1 & +18 34 16 & 0.073 & 0149900301 & 14.0, 13.6, 11.1 & 1.793 & 0.391$\pm$0.047 & 0.023$\pm$0.001 & 0.105$\pm$0.002 \\
S0792 & 17 05 10.3 & -82 10 26 & 0.074 & 0761111801 & 10.7, 11.3, 8.2 & 1.815 & 0.390$\pm$0.044 & 0.004$\pm$0.000 & 0.280$\pm$0.004 \\
A0401 & 02 58 57.5 & +13 34 46 & 0.074 & 0112260301 & 12.2, 11.6, 7.4 & 5.849 & 0.405$\pm$0.028 & 0.010$\pm$0.001 & 0.208$\pm$0.001 \\
  \hline
 \end{tabular}
\end{table*}

\begin{table*}
 \contcaption{}
\begin{tabular}{|c|ccc|ccc|ccc}
\hline
Name & RA & DEC & redshift  & ObsID & $t_{exp}$ & $M_{500}$ & A$_{core}$ & w & c  \\
& & & &  & [ks] & [$10^{14} M_{\odot}$] & [solar] &  & \\
\hline
Z4905+0523 & 12 10 18.8 & +05 23 06 & 0.075 & 0765030501 & 7.3, 7.1, 4.9 & 1.616 & 0.559$\pm$0.082 & 0.010$\pm$0.001 & 0.227$\pm$0.004 \\
A3825 & 21 58 27.2 & -60 23 58 & 0.075 & 0675472201 & 12.4, 12.4, 10.0 & 2.002 & 0.357$\pm$0.071 & 0.009$\pm$0.001 & 0.078$\pm$0.002 \\
A3112 & 03 17 58.5 & -44 14 20 & 0.075 & 0603050101 & 80.6, 83.3, 52.6 & 4.395 & 0.660$\pm$0.006 & 0.011$\pm$0.001 & 0.505$\pm$0.001 \\
& & & & 0603050201 & 66.1, 66.3, 44.9 & & & & \\
A3806 & 21 46 20.9 & -57 17 19 & 0.076 & 0675472101 & 12.6, 12.2, 10.2 & 1.841 & 0.409$\pm$0.036 & 0.011$\pm$0.001 & 0.276$\pm$0.003 \\
A3822 & 21 54 09.2 & -57 51 19 & 0.076 & 0675470401 & 9.5, 10.7, 2.9 & 3.050 & 0.352$\pm$0.031 & 0.018$\pm$0.001 & 0.146$\pm$0.002 \\
& & & & 0720250301 & 8.0, 8.1, 3.4 & & & & \\
A2670 & 23 54 13.4 & -10 24 46 & 0.076 & 0108460301 & 12.2, 12.4, 6.9 & 2.316 & 0.527$\pm$0.040 & 0.008$\pm$0.000 & 0.235$\pm$0.003 \\
Z1215+0349 & 12 17 40.6 & +03 39 45 & 0.077 & 0300211401 & 21.8, 22.1, 13.9 & 3.592 & 0.331$\pm$0.024 & 0.009$\pm$0.001 & 0.158$\pm$0.001 \\
A2029 & 15 10 55.0 & +05 43 12 & 0.077 & 0111270201 & 10.6, 11.0, 8.2 & 7.271 & 0.574$\pm$0.010 & 0.003$\pm$0.000 & 0.465$\pm$0.001 \\
& & & & 0551780201 & 28.7, 31.3, 15.0 & & & & \\
& & & & 0551780301 & 35.5, 36.7, 21.4 & & & & \\
& & & & 0551780401 & 29.2, 30.4, 14.3 & & & & \\
& & & & 0551780501 & 22.8, 23.2, 16.2 & & & & \\
A1648 & 12 58 49.8 & -26 40 03 & 0.077 & 0765040701 & 11.4, 11.7, 9.0 & 1.601 & 0.184$\pm$0.050 & 0.015$\pm$0.001 & 0.127$\pm$0.004 \\
A3638 & 19 25 29.6 & -42 56 57 & 0.077 & 0765020101 & 11.2, 11.7, 7.5 & 1.859 & 0.475$\pm$0.030 & 0.005$\pm$0.000 & 0.395$\pm$0.004 \\
A2061 & 15 21 17.0 & +30 38 24 & 0.078 & 0721740101 & 44.0, 43.4, 35.7 & 2.854 & 0.255$\pm$0.020 & 0.050$\pm$0.003 & 0.084$\pm$0.001 \\
A1205 & 11 13 20.7 & +02 31 56 & 0.078 & 0720250701 & 7.0, 7.2, 4.8 & 2.043 & 0.555$\pm$0.112 & 0.059$\pm$0.003 & 0.106$\pm$0.003 \\
R2344.2--0422 & 23 44 16.0 & -04 22 03 & 0.079 & 0677180501 & 11.6, 11.6, 6.9 & 2.822 & 0.431$\pm$0.035 & 0.010$\pm$0.001 & 0.199$\pm$0.002 \\
A1035 & 10 32 14.8 & +40 15 53 & 0.079 & 0653810501 & 15.1, 15.2, 12.6 & 1.712 & 0.515$\pm$0.050 & 0.009$\pm$0.001 & 0.336$\pm$0.005 \\
R2224.5--5515 & 22 24 27.5 & -55 15 22 & 0.079 & 0765020801 & 10.8, 11.0, 6.0 & 1.772 & 0.387$\pm$0.080 & 0.028$\pm$0.002 & 0.112$\pm$0.003 \\
R0229.3--3332 & 02 29 22.3 & -33 32 16 & 0.079 & 0677180801 & 6.8, 7.4, 3.9 & 1.362 & 0.280$\pm$0.063 & 0.011$\pm$0.001 & 0.382$\pm$0.006 \\
S0700 & 12 36 44.7 & -33 54 10 & 0.080 & 0201903701 & 11.6, 11.9, 7.4 & 1.633 & 0.523$\pm$0.026 & 0.011$\pm$0.001 & 0.280$\pm$0.003 \\ 
& & & & 0302610701 & 21.9, 22.0, 15.9 & & & & \\
A3771 & 21 29 51.0 & -50 48 04 & 0.080 & 0201902501 & 16.5, 17.7, 8.9 & 1.614 & 0.418$\pm$0.041 & 0.042$\pm$0.002 & 0.102$\pm$0.002 \\
& & & & 0654440201 & 24.9, 31.7, 8.3 & & & & \\
A2377 & 21 45 54.8 & -10 06 16 & 0.081 & 0675472001 & 10.8, 11.3, 4.8 & 2.034 & 0.539$\pm$0.058 & 0.007$\pm$0.000 & 0.160$\pm$0.003 \\
A2255 & 17 12 47.2 & +64 03 41 & 0.081 & 0112260801 & 7.6, 7.4, 2.9 & 3.741 & 0.362$\pm$0.062 & 0.104$\pm$0.006 & 0.121$\pm$0.006 \\
A2402 & 21 58 30.5 & -09 48 28 & 0.081 & 0765030101 & 9.1, 9.0, 5.4 & 1.787 & 0.623$\pm$0.055 & 0.008$\pm$0.000 & 0.426$\pm$0.006 \\
A2410 & 22 02 05.9 & -09 49 28 & 0.081 & 0720252101 & 12.5, 12.4, 7.6 & 1.788 & 0.386$\pm$0.069 & 0.015$\pm$0.001 & 0.172$\pm$0.004 \\
A0653 & 08 21 51.7 & +01 12 42 & 0.082 & 0201903601 & 10.3, 9.8, 6.2 & 1.479 & 0.345$\pm$0.047 & 0.020$\pm$0.001 & 0.198$\pm$0.004 \\
& & & & 0404911201 & 13.7, 14.0, 8.8 & & & & \\
A2566 & 23 16 07.5 & -20 27 19 & 0.082 & 0677180301 & 7.1, 8.4, 3.6 & 2.173 & 0.577$\pm$0.027 & 0.003$\pm$0.000 & 0.476$\pm$0.004 \\
R2143.9--5637 & 21 43 58.3 & -56 37 35 & 0.082 & 0675471901 & 12.5, 12.4, 10.2 & 2.650 & 0.443$\pm$0.025 & 0.005$\pm$0.000 & 0.477$\pm$0.004 \\
A2428 & 22 16 15.5 & -09 20 24 & 0.083 & 0675472401 & 10.4, 10.0, 7.9 & 2.412 & 0.467$\pm$0.031 & 0.008$\pm$0.000 & 0.337$\pm$0.003 \\
A2245 & 17 02 31.9 & +33 30 47 & 0.084 & 0672910101 & 12.4, 13.0, 10.0 & 1.242 & 0.604$\pm$0.070 & 0.013$\pm$0.001 & 0.210$\pm$0.005 \\
A1650 & 12 58 41.1 & -01 45 25 & 0.084 & 0093200101 & 32.2, 32.5, 26.9 & 4.121 & 0.456$\pm$0.016 & 0.007$\pm$0.000 & 0.323$\pm$0.001 \\
A1651 & 12 59 21.5 & -04 11 41 & 0.085 & 0203020101 & 7.7, 8.2, 5.8 & 4.393 & 0.517$\pm$0.025 & 0.008$\pm$0.000 & 0.290$\pm$0.002 \\
A2420 & 22 10 19.7 & -12 10 34 & 0.085 & 0675470501 & 11.9, 12.5, 9.0 & 3.514 & 0.481$\pm$0.036 & 0.027$\pm$0.002 & 0.159$\pm$0.002 \\
A1663 & 13 02 50.7 & -02 30 22 & 0.085 & 0201901801 & 21.7, 21.6, 15.0 & 1.605 & 0.703$\pm$0.040 & 0.030$\pm$0.002 & 0.310$\pm$0.003 \\
A1750 & 13 30 49.9 & -01 52 22 & 0.085 & 0112240301 & 25.9, 27.1, 20.8 & 3.057 & 0.412$\pm$0.039 & 0.018$\pm$0.001 & 0.186$\pm$0.002 \\
A2597 & 23 25 20.0 & -12 07 38 & 0.085 & 0723801601 & 20.6, 22.4, 9.2 & 4.182 & 0.413$\pm$0.007 & 0.003$\pm$0.000 & 0.580$\pm$0.002 \\
& & & & 0723801701 & 41.3, 45.2, 20.2 & & & & \\
A3126 & 03 28 37.5 & -55 42 46 & 0.085 & 0675471001 & 5.1, 5.2, 3.0 & 2.437 & 0.539$\pm$0.054 & 0.004$\pm$0.000 & 0.218$\pm$0.004 \\
& & & & 0720250501 & 5.0, 6.0, - & & & & \\
A2556 & 23 13 00.9 & -21 37 55 & 0.087 & 0677180201 & 8.4, 8.5, 4.2 & 2.476 & 0.510$\pm$0.036 & 0.007$\pm$0.000 & 0.460$\pm$0.004 \\
R1309.2-0136 & 13 09 17.0 & -01 36 45 & 0.088 & 0201750201 & 6.1, 6.0, 5.0 & 1.860 & 0.756$\pm$0.103 & 0.010$\pm$0.001 & 0.294$\pm$0.006 \\
A0478 & 04 13 25.6 & +10 28 01 & 0.088 & 0109880101 & 62.1, 67.0, 44.1 & 6.424 & 0.458$\pm$0.007 & 0.003$\pm$0.000 & 0.455$\pm$0.001 \\
A0278 & 01 57 25.7 & +32 13 26 & 0.089 & 0203980101 & 20.4, 21.0, 15.7 & 1.480 & 0.265$\pm$0.039 & 0.012$\pm$0.001 & 0.219$\pm$0.003 \\
A2142 & 15 58 20.6 & +27 13 37 & 0.091 & 0674560201 & 48.7, 48.5, 37.5 & 8.149 & 0.408$\pm$0.010 & 0.009$\pm$0.001 & 0.323$\pm$0.001 \\
A3695 & 20 34 47.9 & -35 48 48 & 0.089 & 0675470301 & 10.4, 10.6, 7.6 & 3.730 & 0.387$\pm$0.045 & 0.019$\pm$0.001 & 0.127$\pm$0.002 \\
A3998 & 23 21 33.4 & -41 53 56 & 0.089 & 0677180401 & 7.4, 8.2, 4.3 & 2.586 & 0.591$\pm$0.041 & 0.005$\pm$0.000 & 0.391$\pm$0.004 \\
A2442 & 22 25 51.0 & -06 36 12 & 0.090 & 0677182201 & 7.6, 8.5, 3.2 & 1.912 & 0.378$\pm$0.044 & 0.006$\pm$0.000 & 0.251$\pm$0.005 \\
& & & & 0677182701 & 7.7, 8.0, 5.7 & & & & \\
R1301.6-0650 & 13 01 36.3 & -06 50 00 & 0.090 & 0677181701 & 11.4, 10.8, 8.1 & 1.588 & 0.519$\pm$0.078 & 0.022$\pm$0.001 & 0.344$\pm$0.008 \\
UGC09480 & 14 42 18.4 & +22 18 17 & 0.090 & 0765010501 & 11.0, 10.8, 6.4 & 2.323 & 0.582$\pm$0.040 & 0.003$\pm$0.000 & 0.448$\pm$0.005 \\
  \hline
 \end{tabular}
\end{table*}

\begin{table*}
 \contcaption{}
\begin{tabular}{|c|ccc|ccc|ccc}
\hline
Name & RA & DEC & redshift  & ObsID & $t_{exp}$ & $M_{500}$ & A$_{core}$ & w & c  \\
& & & &  & [ks] & [$10^{14} M_{\odot}$] & [solar] &  & \\
\hline
Z1703--0132 & 17 06 26.6 & -01 32 23 & 0.091 & 0675471401 & 6.2, 6.4, 4.3 & 2.139 & 0.470$\pm$0.059 & 0.017$\pm$0.001 & 0.128$\pm$0.003 \\
& & & & 0720250901 & 7.7, 7.1, 5.7 & & & & \\
A3358 & 05 38 16.3 & -20 37 23 & 0.091 & 0677181201 & 8.7, 8.7, 7.7 & 1.887 & 0.335$\pm$0.057 & 0.012$\pm$0.001 & 0.157$\pm$0.003 \\
A0761 & 09 10 36.3 & -10 34 52 & 0.092 & 0765040601 & 6.4, 6.7, 3.1 & 1.994 & 0.405$\pm$0.087 & 0.007$\pm$0.000 & 0.167$\pm$0.004 \\
A2700 & 00 03 50.6 & +02 03 48 & 0.092 & 0201900101 & 23.1, 23.4, 15.3 & 1.734 & 0.447$\pm$0.039 & 0.003$\pm$0.000 & 0.245$\pm$0.003 \\
R0548.8-2154 & 05 48 50.4 & -21 54 43 & 0.093 & 0677181301 & 22.5, 23.0, 18.6 & 1.659 & 0.442$\pm$0.048 & 0.048$\pm$0.003 & 0.095$\pm$0.002 \\
A2312 & 18 53 58.1 & +68 22 53 & 0.093 & 0692930701 & 13.8, 13.7, 6.8 & 2.041 & 0.539$\pm$0.059 & 0.042$\pm$0.002 & 0.230$\pm$0.003 \\
A3694 & 20 34 42.1 & -34 04 26 & 0.094 & 0675471701 & 10.4, 10.8, 8.1 & 2.744 & 0.521$\pm$0.040 & 0.023$\pm$0.001 & 0.289$\pm$0.003 \\
A0013 & 00 13 38.3 & -19 30 08 & 0.094 & 0200270101 & 29.5, 29.5, 25.4 & 2.182 & 0.357$\pm$0.037 & 0.023$\pm$0.001 & 0.163$\pm$0.002 \\
A3921 & 22 49 57.0 & -64 25 46 & 0.094 & 0112240101 & 27.6, 26.6, - & 3.614 & 0.372$\pm$0.022 & 0.010$\pm$0.001 & 0.192$\pm$0.001 \\
RBS1847 & 22 18 05.5 & -65 11 06 & 0.095 & 0675470701 & 11.9, 12.6, 9.2 & 2.795 & 0.400$\pm$0.037 & 0.011$\pm$0.001 & 0.272$\pm$0.003 \\
A2244 & 17 02 42.9 & +34 03 43 & 0.095 & 0740900101 & 24.1, 24.1, 18.1 & 4.491 & 0.425$\pm$0.018 & 0.007$\pm$0.000 & 0.349$\pm$0.001 \\
A4010 & 23 31 12.7 & -36 30 24 & 0.096 & 0404520501 & 17.1, 16.7, 12.3 & 3.282 & 0.723$\pm$0.034 & 0.006$\pm$0.000 & 0.411$\pm$0.003 \\
A3911 & 22 46 18.6 & -52 43 46 & 0.097 & 0149670301 & 22.1, 22.4, 16.5 & 3.399 & 0.331$\pm$0.031 & 0.008$\pm$0.001 & 0.110$\pm$0.001 \\
R1558.3--1410 & 15 58 23.2 & -14 10 04 & 0.097 & 0675472901 & 9.6, 10.5, 7.5 & 3.873 & 0.561$\pm$0.015 & 0.005$\pm$0.000 & 0.425$\pm$0.002 \\
& & & & 0720253101 & 29.0, 29.8, 21.8 & & & & \\
A2175 & 16 20 31.7 & +29 53 43 & 0.097 & 0692930901 & 10.4, 10.1, 8.4 & 2.573 & 0.440$\pm$0.057 & 0.009$\pm$0.001 & 0.176$\pm$0.003 \\
R1931.6--3354 & 19 31 38.7 & -33 54 47 & 0.097 & 0675471601 & 9.9, 9.7, 6.5 & 2.381 & 0.451$\pm$0.039 & 0.003$\pm$0.000 & 0.421$\pm$0.005 \\
R1633.8-0738 & 16 33 53.9 & -07 38 42 & 0.097 & 0677181901 & 10.7, 11.3, 6.1 & 1.614 & 0.206$\pm$0.082 & 0.029$\pm$0.002 & 0.130$\pm$0.004 \\
A3827 & 22 01 56.0 & -59 56 58 & 0.098 & 0149670101 & 19.3, 19.1, 14.6 & 4.589 & 0.403$\pm$0.025 & 0.004$\pm$0.000 & 0.233$\pm$0.001 \\
A2426 & 22 14 32.6 & -10 22 18 & 0.098 & 0675470601 & 10.1, 10.1, 7.8 & 3.558 & 0.592$\pm$0.036 & 0.007$\pm$0.000 & 0.313$\pm$0.003 \\
A0550 & 05 52 52.4 & -21 03 25 & 0.099 & 0675470101 & 5.1, 6.6, - & 3.191 & 0.442$\pm$0.032 & 0.025$\pm$0.001 & 0.175$\pm$0.002 \\
& & & & 0720250101 & 15.2, 15.0, 12.7 & & & & \\
A4067 & 23 59 19.2 & -60 42 00 & 0.099 & 0677180601 & 9.9, 10.1, 8.0 & 1.970 & 0.635$\pm$0.089 & 0.099$\pm$0.005 & 0.147$\pm$0.003 \\
  \hline
 \end{tabular}
\end{table*}

\bsp	
\label{lastpage}
\end{document}